\documentclass[sigconf,screen]{acmart}
\acmConference[ICSE 2024]{46th International Conference on Software Engineering}{April 2024}{Lisbon, Portugal}

\usepackage{multirow}
\usepackage{subcaption}
\usepackage{enumitem}
\usepackage{algorithmic}
\usepackage{algorithm}
\usepackage{booktabs}
\usepackage{graphicx} 
\usepackage{bm}
\usepackage{xspace}
\usepackage{caption}
\usepackage{url}
\usepackage{amsmath,amsfonts}
\usepackage{textcomp}
\usepackage{xcolor}
\usepackage{pifont}
\usepackage[most]{tcolorbox}

\setlist[enumerate]{nolistsep}

\newcommand{\tool}{HINT\xspace}

\usepackage{tikz}
\newcommand*{\circled}[1]{\lower.7ex\hbox{\tikz\draw (0pt, 0pt)%
    circle (.5em) node {\makebox[1em][c]{\small #1}};}}
    
\AtBeginDocument{%
  \providecommand\BibTeX{{%
    \normalfont B\kern-0.5em{\scshape i\kern-0.25em b}\kern-0.8em\TeX}}}

\setcopyright{acmcopyright}
\acmPrice{15.00}
\acmYear{2023}
\copyrightyear{2023}
\acmConference[ICSE '24]{Proceedings of the 46th International Conference on Software Engineering}{April 14--20, 2024}{Singapore, Singapore}
\acmBooktitle{Proceedings of the 46th International Conference on Software Engineering (ICSE '24), April 14--20, 2024, Lisbon, Portugal}

\begin{document}
\title{Learning in the Wild: Towards Leveraging Unlabeled Data
for Effectively Tuning Pre-trained Code Models}



\author{Shuzheng Gao}
\affiliation{%
  \institution{The Chinese University of Hong Kong}
  \city{Hong Kong}
  \country{China}}
\email{szgao23@cse.cuhk.edu.hk}

\author{Wenxin Mao}
\affiliation{%
  \institution{Harbin Institute of Technology}
  \city{Shenzhen}
  \country{China}}
\email{maowx5519@mails.jlu.edu.cn}

\author{Cuiyun Gao}
\authornote{Corresponding author. The author is also affiliated with Peng Cheng Laboratory and Guangdong Provincial Key Laboratory of Novel Security Intelligence Technologies. }
\affiliation{%
  \institution{Harbin Institute of Technology}
  \city{Shenzhen}
  \country{China}}
\email{gaocuiyun@hit.edu.cn}

\author{Li Li}
\affiliation{%
  \institution{Beihang university}
  \city{Beijing}
  \country{China}}
\email{lilicoding@ieee.org}

\author{Xing Hu, Xin Xia}
\affiliation{%
  \institution{Zhejiang university}
  \city{Zhejiang}
  \country{China}}
\email{xinghu@zju.edu.cn, xin.xia@acm.org}


\author{Michael R. Lyu}
\affiliation{%
  \institution{The Chinese University of Hong Kong}
  \city{Hong Kong}
  \country{China}}
\email{lyu@cse.cuhk.edu.hk}

\begin{abstract}

Pre-trained code models have recently achieved substantial improvements in many code intelligence tasks. These models are first pre-trained on large-scale unlabeled datasets in a \textit{task-agnostic} manner using self-supervised learning, and then fine-tuned on labeled datasets in downstream tasks. However, the labeled datasets are usually limited in size (i.e., human intensive efforts), which may hinder the performance of pre-trained code models in specific tasks. To mitigate this, one possible solution is to leverage the large-scale unlabeled data in the tuning stage by pseudo-labeling, i.e., generating pseudo labels for unlabeled data and further training the pre-trained code models with the pseudo-labeled data. However, directly employing the pseudo-labeled data can bring a large amount of noise, i.e., incorrect labels, leading to suboptimal performance. How to effectively leverage the noisy pseudo-labeled data is a  challenging yet under-explored problem.

In this paper, we propose a novel approach named \tool to improve pre-trained code models with large-scale unlabeled datasets by better utilizing the pseudo-labeled data.
HINT includes two main modules:
\textbf{H}ybr\textbf{I}d pseudo-labeled data selection and \textbf{N}oise-tolerant \textbf{T}raining. 
In the hybrid pseudo-data selection module, considering the robustness issue, apart from directly measuring the quality of pseudo labels through training loss, we propose to further employ a retrieval-based method to filter low-quality pseudo-labeled data.
The noise-tolerant training module aims to further mitigate the influence of errors in pseudo labels by training the model with a noise-tolerant loss function and by regularizing the consistency of model predictions. We evaluate the effectiveness of \tool on three popular code intelligence tasks, including code summarization, defect detection, and assertion generation. We build our method on top of three popular open-source pre-trained code models. 
The experimental results show that \tool can better leverage those unlabeled data in a \textit{task-specific} way and provide complementary benefits for pre-trained models, e.g.,  improving the 
best baseline model  
by 15.33\%, 16.50\%, and 8.98\% on code summarization, defect detection, and assertion generation, respectively.
\end{abstract}


\pagestyle{plain}

\maketitle

\section{Introduction}\label{sec:intro}

Recently, code intelligence has become a popular research field in software engineering. It aims at improving developers' productivity by providing real-time coding assistance and suggestions for them~\cite{DBLP:journals/corr/abs-2302-04098,DBLP:conf/icse/HuX0WCZ22}. The advent of deep learning techniques, especially pre-training techniques~\cite{radford2018improving,DBLP:conf/naacl/DevlinCLT19}, has significantly advanced progress in this area. Different from previous supervised learning methods that train the model from scratch~\cite{DBLP:conf/acl/AhmadCRC20,DBLP:conf/acl/YinN17}, these pre-trained code models are first pre-trained on large-scale unlabeled datasets using self-supervised learning tasks and then fine-tuned on labeled datasets in downstream tasks. For example, Masked Language Modeling (MLM) is one of the most popular self-supervised pre-training tasks and is used in many pre-trained code models such as CodeBERT~\cite{DBLP:conf/emnlp/FengGTDFGS0LJZ20} and GraphCodeBERT~\cite{DBLP:conf/iclr/GuoRLFT0ZDSFTDC21}. It works by training the models to predict the masked tokens based on the context of
surrounding words. Since this process does not require human annotation, it can be applied on large-scale unlabeled datasets, enabling
the models to acquire a vast amount of general programming knowledge. Equipped with this ability, these pre-trained code models
achieve state-of-the-art performance on a variety of code intelligence tasks, such as code summarization and defect detection~\cite{DBLP:conf/emnlp/FengGTDFGS0LJZ20,DBLP:conf/iclr/GuoRLFT0ZDSFTDC21,DBLP:conf/acl/GuoLDW0022,DBLP:conf/kbse/GaoWGWZL23}.


Despite the promising results, 
deep learning models are known to be data-hungry and the size of labeled datasets in downstream tasks is important for the performance of pre-trained models~\cite{DBLP:journals/aiopen/HanZDGLHQYZZHHJ21,DBLP:journals/csur/WangYKN20}. However, the sizes of labeled datasets in downstream tasks
are usually limited 
due to two main reasons.
On one hand, the datasets crawled from open-source websites like Github or Stackoverflow are small in size and of low quality. For example, as mentioned in the literature~\cite{DBLP:journals/corr/abs-1909-09436}, only 6.8\% JavaScript code snippets from popular GitHub repositories contain corresponding comments, making only a few of them usable for tasks like code summarization. Furthermore, recent studies have revealed that the quality of existing crawled datasets is also quite poor~\cite{DBLP:journals/corr/abs-2207-05579,DBLP:conf/icse/SunLL0022,DBLP:conf/icse/CroftBK23}. For example, as indicated in a recent work~\cite{DBLP:journals/corr/abs-2207-05579}, over 40\% of data in the widely-used code summarization datasets contain various types of noise. On the other hand, due to the requirement of domain expert knowledge, the annotation cost of code intelligence tasks is higher than other tasks in natural language processing or computer vision, such as sentiment analysis and image classification~\cite{MTurk}.
With insufficient annotated data in downstream tasks, the performance of pre-trained code models is limited.

\begin{figure}[t]
    \centering
    \begin{subfigure}[b]{0.45\textwidth}
        \centering
        \includegraphics[width=1\textwidth]{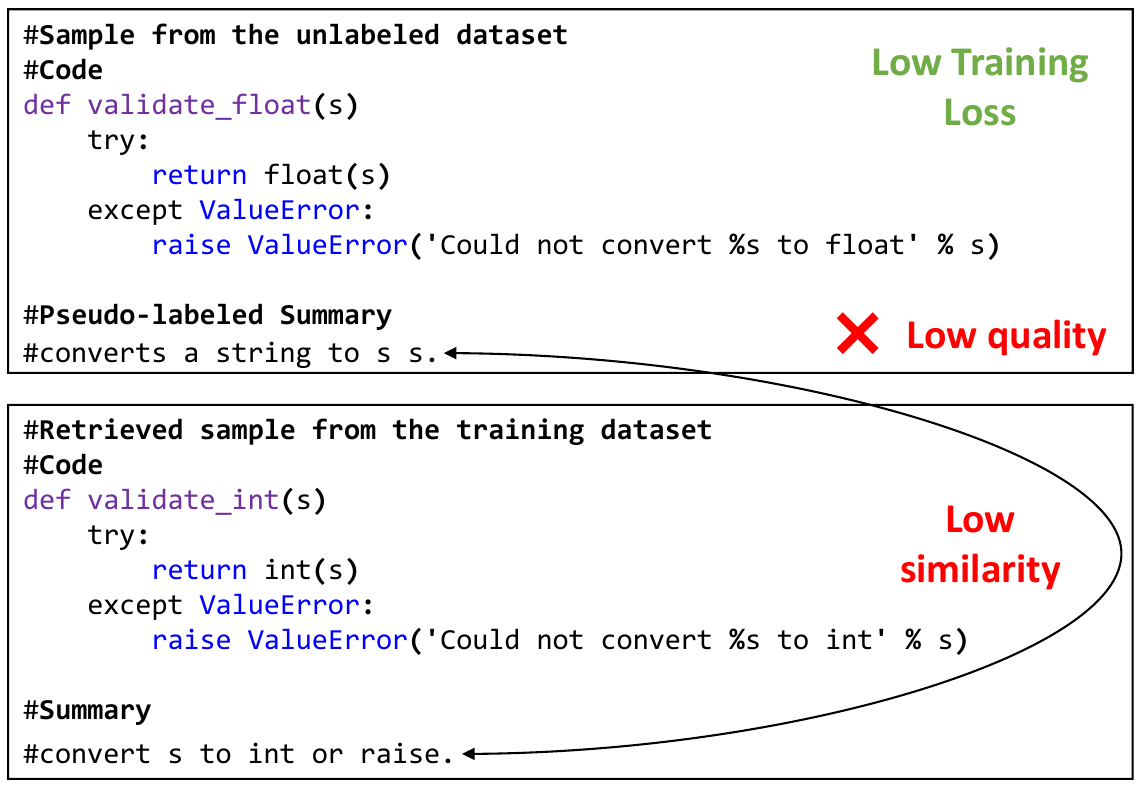}
        \caption{A Python example of low-quality pseudo-labeled data (top).}
        \label{fig:example_1}
      \end{subfigure}
      \hfill
      \begin{subfigure}[b]{0.45\textwidth}
        \centering
        \includegraphics[width=1\textwidth]{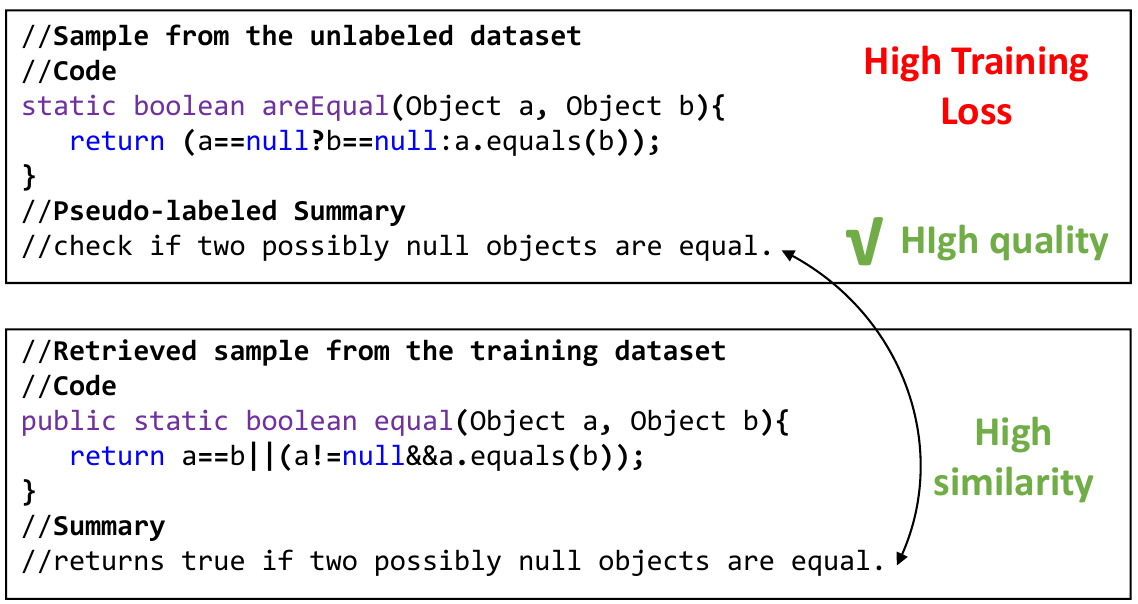}
        \caption{A Java example of high-quality pseudo-labeled data (top). 
        }
        \label{fig:example_2}
      \end{subfigure}
    \caption{Examples in the code summarization task for illustrating the motivation of the hybrid pseudo-labeled data selection method, which indicates the loss-based data selection strategy alone may incorrectly measure the quality of pseudo labels.
    }
    \vspace{-0.4cm}
	\label{fig:motivated_example}
\end{figure}

One possible solution to this problem is to leverage the large-scale unlabeled data in the tuning stage by pseudo-labeling.
Pseudo-labeling first trains a base model on the limited labeled dataset, which subsequently serves as a teacher model to annotate the unlabeled dataset~\cite{lee2013pseudo,DBLP:conf/acl/Jiao0T0LK20,DBLP:conf/emnlp/MiZ0CHF21}. The pseudo-labeled dataset is then merged with the original labeled dataset to help improve the training of a new student model. By replacing the teacher model with the stronger student model, the above process can be iterated 
multiple
times, aiming at improving the models themselves.
This technique leverages the unlabeled data in a \textit{task-specific} way and has shown promising results in tasks such as image classification~\cite{lee2013pseudo}
and dialog systems~\cite{DBLP:conf/emnlp/MiZ0CHF21}. 
Although pseudo-labeling can enrich the labeled dataset, directly employing the pseudo-labeled data can bring a large amount of noise~\cite{DBLP:conf/emnlp/MiZ0CHF21}. For example, as shown in Figure~\ref{fig:motivated_example} (a), the pseudo-labeled summary of the top code snippet is not a meaningful sentence and contains redundant tokens.
Training with such noisy pseudo labels may amplify the incorrect knowledge in the teacher model and ultimately degrades the model's performance. However, identifying and removing noisy pseudo labels is non-trivial due to the complex semantic of source code. Besides, it is difficult and impractical to ensure that the filtered dataset is
noise-free~\cite{DBLP:conf/nips/ZhangS18,DBLP:conf/iccv/0001MCLY019}. 
Therefore, how to effectively leverage the noisy pseudo-labeled data and enable the model to be noise-tolerant for code intelligence tasks is of vital importance, yet under-explored.

In this paper, we propose \tool with two main components, i.e., the \textbf{H}ybr\textbf{I}d pseudo-labeled data selection module and the \textbf{N}oise-tolerant \textbf{T}raining module. 
First, in the hybrid pseudo-labeled data selection module, 
we propose to combine the training loss of the teacher model and a retrieval-based method for removing
the low-quality data. 
Specifically, we filter out pseudo-labeled samples that present
high training loss or low label similarity with the retrieved similar training sample.
To further mitigate the influence of data noise on model performance, we propose a noise-tolerant training objective that includes a noise-tolerant symmetric loss function and a consistency regularization of model predictions. 
To evaluate the performance of \tool, we conduct experiments on three popular code intelligence tasks including code summarization, defect detection, and assertion generation. Following previous work~\cite{DBLP:conf/icse/WangCLLPDL23,DBLP:conf/icse/GaoZGW23,DBLP:conf/icse/NiuLNCGL23}, we build our method on top of three popular open-source pre-trained models: CodeBERT~\cite{DBLP:conf/emnlp/FengGTDFGS0LJZ20}, CodeT5~\cite{DBLP:conf/emnlp/0034WJH21}, and UniXcoder~\cite{DBLP:conf/acl/GuoLDW0022}. 
Extensive experiments demonstrate that \tool can consistently improve the performance of pre-trained code models on these code intelligence tasks. For example, \tool improves UniXcoder by 15.33\%, 16.50\%, and 8.98\% in terms of BLEU-4, F1, and EM on code summarization, defect detection, and assertion generation, respectively, indicating that our proposed \tool method can provide complementary benefits for the pre-trained code models.

\begin{figure*}
    \centering
    \includegraphics[width=0.85\textwidth]{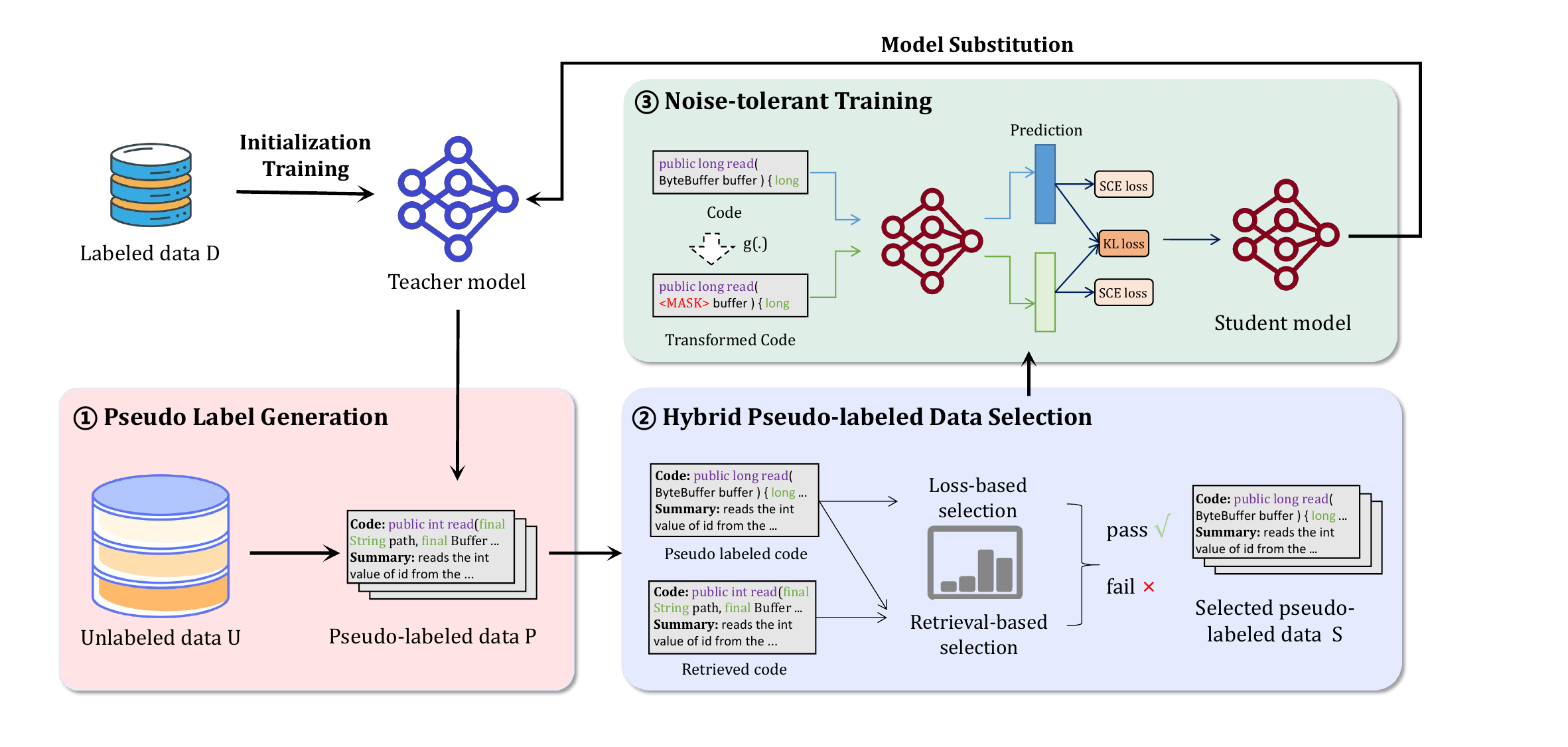}
    \caption{The overview of \tool.}
    \label{fig:approach}
    \vspace{-0.35cm}
\end{figure*}

In summary, the main contributions of this work are as follows:
\begin{enumerate}
    \item To the best of our knowledge, we are the first to leverage the large-scale unlabeled data in a task-specific way in the turning phase for code intelligence tasks.
    \item We propose \tool, a novel framework to leverage large-scale unlabeled data for effectively tuning pre-trained code models. It first selects high-quality pseudo-labeled data in a hybrid way and then improves the model's tolerance to noisy data in the training process.  
    \item Extensive experiments on three tasks demonstrate that our method can be built on top of a range of existing strong pre-trained models and consistently improve their performance on many downstream tasks.
\end{enumerate}

\section{Proposed Approach}\label{sec:back}
\subsection{Problem Setup and Overview}
In this section, we explicate the detailed design of \tool. Formally, in code intelligence tasks such as code summarization, we have a set of source codes $X$ and summaries $Y$. Let $D = \{(x^i, y^i)\}^N_{i=1}$ denotes the labeled training dataset, where $x^i \in X , y^i \in Y$ and $N$ denotes the size of $D$. Let $U = \{x^i\}^M_{i=1}$ denote 
the large unlabeled dataset, where $M$ denotes the size of $U$ and $M>N$ in general. Our goal is to learn a model $f:X \mapsto Y$ from both $D$ and $U$ that can well predict the label of input $x^i$ in the test set. 

The overall framework of \tool is shown in Figure~\ref{fig:approach}. We first train a teacher model on the original labeled dataset $D$ and \circled{1}use the teacher model
to generate pseudo labels for the unlabeled dataset. Then, 
\circled{2}a hybrid pseudo-labeled data selection method that contains \textit{loss-based selection} and \textit{retrieval-based selection} is proposed to filter the code with low-quality pseudo labels
(introduced in Section~\ref{subsec:selection}). For further mitigating the influence of noise in pseudo labels during model training, we propose \circled{3}a noise-tolerant training strategy that trains the student model with
noise-tolerant symmetric cross entropy loss and consistency regularization (introduced in Section~\ref{subsec:aug}). The above procedure can be iterated multiple times, 
enabling the models to be self-improved
(introduced in Section~\ref{subsec:ite}). The algorithm is shown in Algorithm~\ref{alg:framework}.

\subsection{Hybrid Pseudo-labeled Data Selection}\label{subsec:selection}
Once we get a trained teacher model $\mathcal{F}_t$, we use it to generate pseudo labels for unlabeled dataset $U$, producing
a pseudo labeled dataset $P = \{(x^i, \hat y^i)\}^M_{i=1}$. The pseudo-labeled data cannot be employed directly, since they may contain substantial noise and impact the model performance. Previous studies in machine learning~\cite{DBLP:conf/nips/HanYYNXHTS18,DBLP:conf/iccv/HuangQJZ19} mainly employ \textit{loss-based selection} by filtering the data with high training loss based on the insight that neural models can well distinguish the quality of each sample (i.e., noisy data are generally associated with higher training loss).
However, code intelligence models are known to suffer from the robustness issue~\cite{DBLP:conf/wcre/HenkelRWAJR22}, so solely relying on the model training loss for noise filtering is ineffective. 
For the example in Figure~\ref{fig:motivated_example} (a), we can observe that although the quality of this generated summary is pretty poor, its loss is low in value. Specifically, when comparing the loss of all the pseudo-labeled data, it exhibits a lower loss than 83\% of the
pseudo-labeled data.
Besides, in Figure~\ref{fig:motivated_example} (b), the generated pseudo summary can well describe the meaning of checking the equivalence of two objects in the Java code snippet but its training loss value is relatively
high, i.e., surpassing 
52\% of the 
pseudo-labeled data. 

Considering that code reuse is widespread in software development~\cite{DBLP:journals/tse/KamiyaKI02,DBLP:conf/sigsoft/KimSN05}, apart from the \textit{loss-based selection}, 
we propose to further select high-quality data through a retrieval-based method. As shown in Figure~\ref{fig:motivated_example}, by comparing the pseudo-labeled summaries and retrieved summaries, we can systematically 
identify the pseudo-labeled data in Figure~\ref{fig:motivated_example} (b) as a high-quality sample and filter the low-quality pseudo-labeled data
in Figure~\ref{fig:motivated_example} (a).
Specifically, in the \textit{retrieval-based selection}, for each unlabeled data $x^i$, 
we first use the widely-used BM-25 method~\cite{manning2009introduction} to retrieve the most similar code $x^j$ 
in the labeled training set. Then we propose to compare the similarities
of $x^i$ and $x^j$ and their corresponding pseudo label $\hat y^i$ and groud truth label $y^j$ through normalized edit distance:

\begin{equation}
NED(x,y)=\left\{\begin{array}{cl}
  \frac{edit\_distance(x,y)}{\left | \left | x  \right | \right |}  & \textbf{if} \ x, y \in sequence\\  
  \text{I} \{x \ne y\} & \textbf{if} \ x, y \in label\\ 
\end{array}\right. 
\end{equation}
where $\left | \left | .  \right | \right |$ denotes the length of the sequence and I$\{ . \}$ is an indicator function that returns 1 if the condition is true and 0 otherwise. Specifically, if both $NED(x^i,x^j)$ and $NED(\hat y^i,y^j)$ are not higher than the threshold $t$, we consider this sample  $(x^i, \hat y^i)$ as a correctly predicted sample and add it to the selected dataset $S$. 
On the contrary, if $NED(x^i,x^j)$ is lower than $t$ while $NED(\hat y^i,y^j)$ is above $1-t$, we choose to filter it as it has a higher probability of being a noisy data (Line 9-11 in Algorithm~\ref{alg:framework}). Here $t$ is a hyperparameter to control the filtering
threshold. 
For samples that cannot be decided by the \textit{retrieval-based selection}, 
we first calculate the training loss of each pseudo-labeled data by re-feeding each code $x^i$ into the teacher model $\mathcal{F}_t$ and using the generated pseudo label $\hat y^i$ as the ground-truth label. We then select the top $K$\% data with the lowest loss values
among all pseudo-labeled data and add them to $S$ (Line 4-5, 13-14 in Algorithm~\ref{alg:framework}). Finally, we obtain a dataset $S = {(x^i, \hat y^i)}_{i=1}^{M'}$ containing high-quality pseudo-labeled data. The dataset $S$ is employed to train the student model, together with the labeled dataset $D$.


\begin{algorithm}[t]
\caption{Algorithm of \tool}
\label{alg:framework}
\begin{algorithmic}[1]
\REQUIRE labeled dataset $D$, unlabeled dataset $U$, threshold of edit distance $t$, the threshold in loss-based selection $K$, code transformation function $g$, iteration number $I$\\
\ENSURE neural model $\mathcal{F}$\\
\STATE Train the teacher model $\mathcal{F}_t$ on $D$
\FOR{each $i$ in $I$}
\STATE Generate pseudo data $P$ for $U$ using $\mathcal{F}_t$ 
\STATE Calculate the loss of samples in $P$ using $\mathcal{F}_t$ 
\STATE TK $\gets$ samples with the least top $K$\% loss value in $P$
\STATE $S\gets\varnothing$ 
\FOR{each sample $\{x_{i}, y'_{i}\}$ in $P$}
\STATE Retrieve the most similar sample $(x_{j}, y_{j})$ from $D$
\IF{$NED(x_{i}, x_{j})\leq t \land NED(y_{i}, y'_{j})\leq t$}
\STATE $S$.\textit{insert}$(\{x_{i}, y'_{i}\})$
\ELSIF{$NED(x_{i}, x_{j})\leq t \land NED(y_{i}, y'_{j})\geq$ $1-t$}
\STATE \textbf{continue}
\ELSIF{$\mathcal{F}_t(x_{i}, y'_{i}) \in$ TK}
\STATE $S$.\textit{insert}$(\{x_{i}, y'_{i}\})$
\ENDIF
\ENDFOR
\STATE $L\gets D \cup S$
\STATE Train the student model $\mathcal{F}_s$ with dataset $L$ and transformation function $g$ through Equation~\ref{equ:robust}
\STATE $\mathcal{F}_t \gets \mathcal{F}_s$ 
\ENDFOR
\RETURN model $\mathcal{F}_t$
\end{algorithmic}
\end{algorithm}

\subsection{Noise-tolerant Training}\label{subsec:aug}
\begin{figure}[t]
    \centering
    \includegraphics[width=0.465\textwidth]{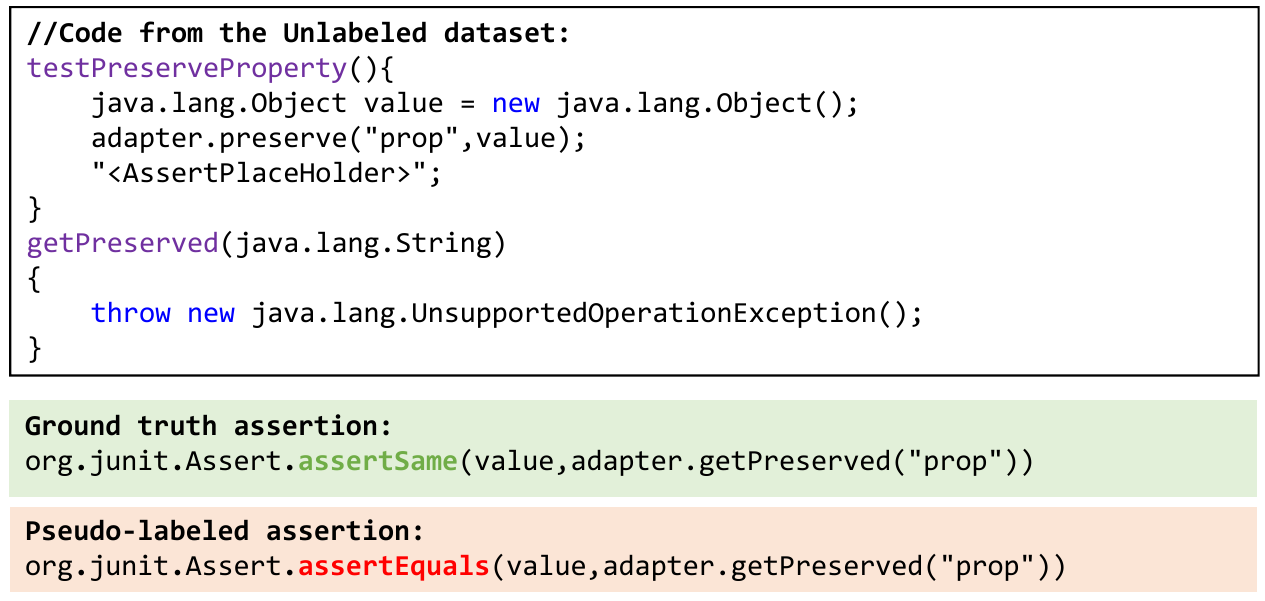}
    \caption{An example of a pseudo label with a minor error.}
    \vspace{-0.45cm}
	\label{fig:noise_useful}
\end{figure}

Despite the dedicated data selection effort, 
it is still difficult and impractical to ensure that the
selected samples $S$ are noise-free. 
Besides, the pseudo-labeled samples with minor noise are not completely harmful and can also provide rich information for model training. For example, as shown in Figure~\ref{fig:noise_useful}, the assertion statement generated by the teacher model mistakenly predicts the assertion type as ``\textit{assertionEquals}''. 
If we directly use it as ground truth to train the model, the model may be misled.
Nevertheless, this sample still contains much valuable information since the predictions on other positions such as the parameters are correct. Therefore, instead of
directly discarding these samples with a more strict filtering process, we propose to leverage the pseudo-labeled data with
noise-tolerant loss function and consistency regularization during model training. 

\textbf{Noise-tolerant Loss Function:} 
Previous studies~\cite{DBLP:conf/nips/ZhangS18,DBLP:conf/iccv/0001MCLY019} have found that the widely-used cross entropy (CE) loss function is sensitive to noisy training data.
Specifically, the coefficient in the gradient of the CE loss function $-\frac{1}{f_{y^i}(x^i;\theta)}\nabla f_{y^i}(x^i;\theta)$ assigns larger weights to samples with higher loss and smaller weights to samples with lower loss. 
Since noisy data often obtain a high training loss in the training process~\cite{DBLP:conf/nips/HanYYNXHTS18,DBLP:conf/iccv/HuangQJZ19}, models trained with CE loss easily focus on those noisy data and tend to be misled.
However, the re-weighting coefficient in the gradient of CE is also beneficial for model training. Directly removing it as $-\nabla f_{y^i}(x^i;\theta)$ might bring the slow convergence problem~\cite{DBLP:conf/nips/ZhangS18}.
To deal with this problem, we propose to employ the Symmetric Cross Entropy loss (SCE)~\cite{DBLP:conf/iccv/0001MCLY019}: 
\begin{equation}\label{equ:sce}
l_{sce}(x,y) = - \sum_{c=1}^{C} (p(c|x_i)\log(q(c|x)) + q(c|x_i)\log(p(c|x))),
\end{equation}
where $q(c|x)$ is the prediction of the model and $p(c|x)$ is the corresponding ground truth label in the dataset. $C$ denotes the number of classes in the classification task or the size of vocabulary in the generation task.  The former item represents the CE loss, while the latter item corresponds to the Reverse Cross Entropy (RCE)~\cite{DBLP:conf/iccv/0001MCLY019} which assigns the same weights for all samples, i.e., $-\nabla f_{y^i}(x^i;\theta)$. {Note that $\log p(c|x)$ is 0 for the ground truth class and the same negative value for other classes. Based on the relation $p(c^*|x) = 1-\sum_{c=1,c \ne c^*}^{C} p(c|x)$ where $c^*$ denotes the ground class, we can achieve the above result.}
In this way, the student model is less likely to be influenced by minor errors in pseudo labels.

\textbf{Consistency Regularization:} 
According to Equation~\ref{equ:sce}, the noise-tolerant loss relies on the quality of pseudo labels. Considering that such supervision signals from pseudo labels might be noisy and unreliable, 
we further propose to add consistency regularization between predictions of the original code and the transformed code. It could enrich the supervision signals and provide the student model with a more reliable objective function
without manual labels.
Specifically, given a code snippet $x$, we first apply a code transformation function $ct(\cdot)$ and obtain the transformed input $ct(x)$. Then, 
the consistency regularization is applied to align the distributions of the predictions $\mathcal{F}_s(x)$ and $\mathcal{F}_s(ct(x))$:


\begin{equation}\label{equ:kl}
\begin{split}
l_{cr} = KL(\mathcal{F}_s(x)||\mathcal{F}_s(ct(x))) + KL(\mathcal{F}_s(ct(x))||\mathcal{F}_s(x)),
\end{split}
\end{equation}
where $KL(\cdot||\cdot)$ denotes the Kullback-Leibler Divergence~\cite{kullback1997information}. {For generation tasks, we approximate it by the average value per token.} For code transformation methods, we follow previous work~\cite{DBLP:journals/corr/abs-2204-03293} and employ four effective transformation methods, 
including Dynamic Masking, Dynamic Replacement, Dynamic Replacement of Specified Type, and Dynamic Masking of Specified Type. For each sample, we randomly select one transformation function in each epoch. 
Different from previous task-agnostic contrastive learning pre-training methods that only focus on aligning the representation of code and transformed code~\cite{DBLP:conf/emnlp/0001JZA0S21,DBLP:conf/acl/GuoLDW0022}, our method directly constrains the prediction $\mathcal{F}_s(x)$ and $\mathcal{F}_s(ct(x))$ of code and transformed code on downstream tasks and regularizes them in a task-specific way.

Finally, \tool trains the student model by combining the noise-tolerant loss function and consistency regularization as follows:

\begin{equation}\label{equ:robust}
\begin{split}
l = \sum_{(x,y) \in D \cup S}^{ } [ \ l_{sce}(x,y) + l_{sce}(ct(x),y) + \mu \cdot l_{cr} \ ],
\end{split}
\end{equation}
where $\mu$ is a hyperparameter to balance the training signals from pseudo labels and the consistency regularization.


\subsection{Iterative Training}\label{subsec:ite}
\begin{sloppypar}
Based on the aforementioned process, we can obtain a student model that has better performance than the teacher model. Then, we can build upon this student model and repeat the process described above to further boost the models themselves (\circled{1}$\to$\circled{2}$\to$\circled{3}$\to$\circled{1}). Specifically, at the end of each iteration, the student model substitutes the teacher model, which is then employed to generate pseudo labels for the unlabeled dataset $D$ in the subsequent iteration.
In general, the better the base model,
the higher the quality of the pseudo labels.
In this way, the student model in the next iteration is more likely to be trained on pseudo-labeled data with higher quality and thus achieves better performance. We follow previous work~\cite{DBLP:conf/emnlp/MiZ0CHF21,DBLP:conf/emnlp/MiCZHF20} and reinitialize the new student model from the pre-trained code models in every iteration. After all iterations, the student model in the last iteration will be used as the final model for predictions on the test set. 
\end{sloppypar}
\section{EXPERIMENTAL setup}\label{sec:setup}

\subsection{Research Questions}
As we claimed above, \tool is a generic framework that works without imposing specific assumptions regarding data distribution, underlying models, or task characteristics, except for the requirement that the input needs to be
in the form of code. To validate the generalizability of \tool,
we propose to evaluate the performance of \tool on a variety of pre-trained code models and three semi-supervised code intelligence tasks with code as input. As for the data distribution, we also evaluate the performance of \tool on cross-domain scenarios that do not have sufficient training data and may present
data distribution gap between training and unlabeled data. Furthermore, we also explore the effectiveness of each component in \tool and the influence of hyperparameters on its performance.
In summary, we evaluate \tool by addressing the following four research questions:

\begin{enumerate}[label=\bfseries RQ\arabic*:,leftmargin=.5in]
    \item How much improvement can \tool provide to existing pre-trained code models?
    \item What is the impact of each component on the performance of \tool?
    \item How well does \tool perform in cross-domain scenarios?
    \item How does \tool's performance vary under different parameter settings?
\end{enumerate}

\subsection{Evaluation Tasks}



We conduct experiments on three representative code intelligence tasks: code summarization, defect detection, and assertion generation, for covering different task types, i.e.,
Code $\to $ Text, Code $\to $ Label, and Code $\to $ Code. 
{Due to the space limitation, we provide a more detailed description of evaluation metrics and statistics of the benchmark datasets in our replication packages~\cite{HINT}.}

\subsubsection{Code Summarization} Code Summarization aims to generate useful comments for a given code snippet. It can help alleviate the developers' cognitive efforts in comprehending programs~\cite{DBLP:journals/infsof/GarousiGRZMS15,DBLP:journals/jss/ChenH09}.

\textbf{Datasets.} In this study, we conduct experiments on two popular benchmark datasets JCSD and PCSD, which contain Java and Python source code, respectively. The JCSD dataset we used is publicly released by Hu et al.~\cite{DBLP:conf/ijcai/HuLXLLJ18}, which contains 87,136 pairs of Java methods and comments collected from 9,714 GitHub repositories. The PCSD dataset comprises 92,545 functions with their respective documentation, which is originally collected by Barone et al.~\cite{barone2017parallel} and later processed by Wei et al.~\cite{DBLP:conf/kbse/Wei19}. For our experiments, we directly used the benchmark datasets released by previous studies~\cite{DBLP:conf/ijcai/HuLXLLJ18,DBLP:conf/acl/AhmadCRC20}, in which the datasets are divided into training, validation, and test sets in a ratio of $8:1:1$ and $6:2:2$ for Java and Python, respectively. As reported in previous work~\cite{DBLP:conf/icse/ShiWD0H00S22,DBLP:conf/kbse/MuC0WW22}, there are duplicated data in the training and test set of the JCSD dataset. 
Therefore, following them, we remove the test samples that also appear in the training or validation set and finally get a deduplicated test set with 6,489 samples. Since there has been no dataset for the evaluation of code intelligence tasks in a semi-supervised setting, we propose to simulate it by extending existing datasets. Specifically, following previous studies~\cite{DBLP:conf/emnlp/MiZ0CHF21,DBLP:conf/ijcai/KeJYHFZH22}, we randomly dividing the initial training data into two subsets: labeled training data and an unlabeled dataset, with the ratio of 9:1.

\textbf{Metrics.} For code summarization, we follow previous work~\cite{DBLP:conf/acl/AhmadCRC20,DBLP:conf/kbse/MuC0WW22,DBLP:journals/tosem/GaoGHZNXL23} and use four popular metrics BLEU-4~\cite{DBLP:conf/acl/PapineniRWZ02}, ROUGE-L~\cite{lin-2004-rouge}, METEOR~\cite{DBLP:conf/acl/BanerjeeL05}, and CIDEr~\cite{DBLP:conf/cvpr/VedantamZP15} for evaluation.

\begin{table*}[t]
\centering
\caption{Experimental results on code summarization. ``*'' denotes statistical significance in comparison to the base models (i.e., two-sided $t$-test with $p$-value$<0.01$).}\label{tab:sum}
\aboverulesep=0ex
\belowrulesep=0ex
\scalebox{1}{
\begin{tabular}{c|c|cccc|cccc}
\toprule
\multicolumn{2}{c|}{\multirow{2}{*}{\textbf{Approach}}} & \multicolumn{4}{c|}{\textbf{JCSD}} & \multicolumn{4}{c}{\textbf{PCSD}}\\
\cmidrule{3-10}
\multicolumn{1}{c}{} & \multicolumn{1}{c|}{} & BLEU-4 & ROUGE-L & METEOR  & CIDEr & BLEU-4 & ROUGE-L &  METEOR & CIDEr \\
\midrule
\multirow{3}{*}{\textbf{CodeBERT}} &  {Base model}  & 13.30 & 26.75 & { 8.10} & 0.58 & 17.94 & 32.35 & { 9.79}  & 0.59\\
& {+HINT\textsubscript{(1)}}  & 14.58* &  \textbf{29.06}* & { 8.74}* & 0.69* & 18.81* & 34.18* & 10.52* & 0.69*\\
&  {+HINT\textsubscript{(5)}}  & \textbf{14.64}* & 29.00* & \textbf{8.87}* & \textbf{0.71}* & \textbf{18.86}* & \textbf{34.25}* & \textbf{10.87}* & \textbf{0.72}*\\
\midrule
\multirow{3}{*}{\textbf{CodeT5}} &  {Base model}  & 16.67 &34.28 & 11.39  & 1.05 & 21.13 & 40.27 & 15.69 & 1.22\\
&  {+HINT\textsubscript{(1)}} & 18.32* &35.49* &  \textbf{12.36}* & 1.22* & 22.33* & 41.42* & \textbf{16.31}* & 1.35*\\
&  {+HINT\textsubscript{(5)}} & \textbf{18.48}* & \textbf{35.63}* &12.29* & \textbf{1.24}* & \textbf{22.55}* & \textbf{41.67}* & 16.21* & \textbf{1.36}*\\
\midrule
\multirow{3}{*}{\textbf{UniXcoder}} &  {Base model}  & 17.16 & 32.56 & 11.05 & 1.11 & 22.42 & 35.84 & 15.38 & 1.31\\
&  {+HINT\textsubscript{(1)}}  & 18.90* & 35.16* & 12.38* & 1.28* & 23.77* & 41.67*& 16.64* & 1.48*\\
&  {+HINT\textsubscript{(5)}}  & \textbf{19.79}* & \textbf{35.83}* & \textbf{13.12}* & \textbf{1.36}* & \textbf{23.98}* & \textbf{41.93}* & \textbf{16.83}* & \textbf{1.50}*\\
\bottomrule
\end{tabular}
}
\end{table*}

\subsubsection{Defect Detection} Defect detection aims at identifying the vulnerabilities in the given program, which is crucial to defend a software system from cyberattack~\cite{fan,DBLP:conf/nips/ZhouLSD019}. 

\textbf{Datasets.} In our experiments, we utilize the widely-used Big-Vul dataset created by Fan et al.~\cite{fan}. This dataset contains 188,636 C/C++ code snippets sourced from more than 300 GitHub projects dating from 2002 to 2019 in Common Vulnerabilities and Exposures (CVE) database. Following previous studies~\cite{fan}, we partition the dataset into training, validation, and test sets with a ratio of 8:1:1. 
Same with code summarization, we also further construct the labeled training data and unlabeled data by dividing the original training set of Big-Vul with a ratio of 1:9.

\textbf{Metrics.}
We follow previous work~\cite{DBLP:conf/nips/ZhouLSD019,DBLP:conf/sigsoft/Li0N21} and evaluate the results by Precision~(P), Recall~(R), and F1. 

\subsubsection{Assertion Generation} Assertion Generation is the task of automatically generating meaningful assert statements for unit tests. It can reduce the manual efforts in writing test cases and facilitate faster detection and diagnosis of software failures~\cite{DBLP:conf/icse/YuLSR00L0W22,mastropaolo2021studying}.

\textbf{Datasets.} 
For assertion generation, we follow previous work~\cite{DBLP:conf/icse/YuLSR00L0W22,mastropaolo2021studying} and use the ATLAS dataset~\cite{watson2020learning}. 
It contains 188,154 real-world test assertions obtained from open-source projects in GitHub. The dataset is composed of eight categories of assertions, and each sample in ATLAS is comprised of a focal method and a test method which serve as the context for generating a single assertion for the given test method. We use the original partition of ATLAS and split it into three subsets: training, validation, and test, in an 8:1:1 ratio. 
The construction of an unlabeled dataset for assertion generation is also the same as the above two tasks. We randomly extract 90\% of the training data for the construction of the unlabeled dataset and use the remaining data as the labeled dataset.

\textbf{Metrics.}
We follow previous work~\cite{DBLP:conf/icse/YuLSR00L0W22,nashidretrieval} in this field and use Exact Match (EM), Longest Common Subsequence (LCS), and Edit Distance (ED) as evaluation metrics. 


\subsection{Baselines}
We evaluate the performance of \tool by building it on the top of three popular open-source pre-trained code models, namely CodeBERT~\cite{DBLP:conf/emnlp/FengGTDFGS0LJZ20}, CodeT5~\cite{DBLP:conf/emnlp/0034WJH21}, and UniXcoder~\cite{DBLP:conf/acl/GuoLDW0022}. \textbf{CodeBERT} is a representative pre-trained code model that is pre-trained with six programming languages and uses Masked Language Modeling and Replace Token Detection as pre-trained tasks. \textbf{CodeT5} is a sequence-to-sequence pre-trained model which involves two code-related pre-training objectives: identifier tagging and masked identifier prediction. It achieves state-of-the-art performance in many sequence generation tasks. \textbf{UniXcoder} is a unified cross-modal pre-trained model which incorporates code semantic and syntax information from AST. It is pre-trained with two new pre-training tasks multi-modal contrastive learning and cross-modal generation to learn code fragment representation. These models are all pre-trained on CodeSearchNet~\cite{DBLP:journals/corr/abs-1909-09436} and CodeT5 is also pre-trained with C/CSharp code snippets from BigQuery~\cite{BigQuery}.

\subsection{Implementation Details}
We reproduce the results of all pre-trained models based on the official repositories
released by the model authors. In order to facilitate a fair comparison, we ensure that the hyperparameters such as training epochs and learning rates for the models with and without \tool are exactly the same. 
In our experiments, we set $\mu$ and $t$ to 0.5 and 0.4, respectively. The maximum iteration is set to five. To determine the percentage of selected samples, we tune the threshold $K$ in 10, 15, 20, 25, 30, or 35 and select the best results for different datasets. Our rationale for hyperparameter selection is discussed in Section~\ref{sec:RQ3}.
When applying our pseudo-labeled data selection methods to the classification task, we conduct Algorithm~\ref{alg:framework} for each class respectively and balance the class distribution by random down-sampling~\cite{bishop2006pattern}.
All the experiments are conducted on an Ubuntu 20.04 server with an Intel Xeon Platinum 8276 CPU, and 4 Nvidia Tesla A100 GPUs which have 40 GB graphic memory.

\section{Experimental Results}\label{sec:result}

\subsection{RQ1: Performance Evaluation}
 In this section, we evaluate the effectiveness of \tool on three code intelligence tasks including code summarization, defect detection, and assertion generation. We present the results of \tool on the first iteration and the best results of \tool on all the five iterations, namely HINT\textsubscript{(1)} and HINT\textsubscript{(5)}. The results are displayed in Table~\ref{tab:sum}-\ref{tab:assert}. \tool consistently improves three pre-train code models on all the tasks and metrics. In particular, \tool achieves 15.33\%, 16.50\%, and 8.98\% improvements in BLEU-4, F1, and EM over the best pre-trained model UniXcoder on the three datasets, respectively. We
 detail the results on each task respectively as below.

\textbf{Code Summarization.} As shown in Table~\ref{tab:sum}, \tool can significantly improve the performance of different existing pre-trained code models on all datasets and metrics even with only one iteration. For example, HINT\textsubscript{(1)} improves the BLEU-4 score of CodeT5 by 9.90\% and 5.68\% on two datasets, respectively. Meanwhile, compared with
the most powerful pre-trained model UniXcoder, HINT\textsubscript{(1)} can still achieve consistent improvement,
e.g., improving UniXcoder on JCSD dataset by 10.14\%, 12.04\%, 7.99\%, and 15.32\% with respect to BLEU-4, METEOR, ROUGE-L, and CIDEr, respectively. This indicates that \tool is effective in leveraging
the unlabeled data and benefits
the strong task-agnostic pre-trained code models in the downstream tasks.

\begin{table}[t]
    \centering
    \caption{Experimental results on defect detection. Statistical significance is not applicable to these metrics
    \cite{conover1999practical}.}
\aboverulesep=0ex
\belowrulesep=0ex
 \scalebox{1}{
    \begin{tabular}{c|c|ccc}
    \toprule
    \multicolumn{2}{c|}{\textbf{Approach}} & Precision & Recall & F1 \\
    \midrule
    \multirow{3}{*}{\textbf{CodeBERT}} &  {Base model}  & 29.64 & 17.63 & 22.11\\
    &  {+HINT\textsubscript{(1)}} & 30.81 & 21.52 & 25.34 \\
    &  {+HINT\textsubscript{(5)}} & \textbf{32.09} & \textbf{22.36} & \textbf{26.35} \\
    \midrule
    \multirow{3}{*}{\textbf{CodeT5}} &  {Base model}  & 31.38 & 20.32 & 24.66\\
    &  {+HINT\textsubscript{(1)}} & 36.79 & 22.36 & 27.81 \\
    &  {+HINT\textsubscript{(5)}} & \textbf{37.66} & \textbf{22.36} & \textbf{28.06} \\
    \midrule
    \multirow{3}{*}{\textbf{UniXcoder}} &  {Base model}  & 31.30 & 17.63 & 22.55\\
    &  {+HINT\textsubscript{(1)}} & \textbf{33.28} & 20.96 & 25.73 \\
    &  {+HINT\textsubscript{(5)}} & 32.04 & \textbf{22.26} & \textbf{26.27} \\  
    \bottomrule
    \end{tabular}
    \label{tab:defect}}
\end{table}

\begin{table}[t]
    \centering
    \caption{Experimental results on assertion generation. ``*'' denotes statistical significance in comparison to the base models (i.e., two-sided $t$-test with $p$-value$<0.01$).}
\aboverulesep=0ex
\belowrulesep=0ex
 \scalebox{1}{
    \begin{tabular}{c|c|ccc}
    \toprule
    \multicolumn{2}{c|}{\textbf{Approach}} & EM & LCS & ED \\
    \midrule
    \multirow{3}{*}{\textbf{CodeBERT}} &  {Base model}  & 31.82 & 65.99 & 21.68\\
    &  {+HINT\textsubscript{(1)}} & 37.75* & 69.46* & \textbf{19.05}* \\
    &  {+HINT\textsubscript{(5)}} & \textbf{38.58}* & \textbf{69.48}* & 19.20* \\
    \midrule
    \multirow{3}{*}{\textbf{CodeT5}} &  {Base model}  & 43.64 & 72.56 & 20.30\\
    &  {+HINT\textsubscript{(1)}} & 46.53* & 74.32* & 18.47* \\
    &  {+HINT\textsubscript{(5)}} & \textbf{47.66}* & \textbf{75.22}* & \textbf{18.17}* \\
    \midrule
    \multirow{3}{*}{\textbf{UniXcoder}} &  {Base model}  & 43.64 & 72.67 & 17.82\\
    &  {+HINT\textsubscript{(1)}} & 47.13* & 74.72* & 16.61* \\
    &  {+HINT\textsubscript{(5)}} & \textbf{47.56}* & \textbf{74.76}* & \textbf{16.21}* \\
    \bottomrule
    \end{tabular}
    \label{tab:assert}}
\end{table}

\begin{table*}[t]
    \centering
    \caption{Ablation study of \tool. Best and second best results are marked in \textbf{bold} and \underline{underline} respectively.}
\aboverulesep=0ex
\belowrulesep=0ex
 \scalebox{1}{
    \begin{tabular}{c|cccc|ccc|ccc}
    \toprule
    \multicolumn{1}{c|}{{\multirow{2}{*}{\textbf{Approach}}}} & \multicolumn{4}{c|}{Code Summarization} &  \multicolumn{3}{c|}{Defect Detection} & \multicolumn{3}{c}{Assertion Generation} \\
    \cmidrule{2-11} 
     &BLEU-4 & ROUGE-L & METEOR & CIDEr & Precision & Recall & F1 & EM  & LCS & ED \\
    \midrule
     \multicolumn{1}{l|}{UniXcoder+\tool} & \underline{18.90} & \textbf{35.16} & \underline{12.38} & \underline{1.28} & 33.28 & \textbf{20.96} & \textbf{25.73} & \textbf{47.13} & \textbf{74.72} & \underline{16.61}\\
    \midrule
    \multicolumn{1}{l|}{Random selection} & 18.34 & 34.11 & 11.70 & 1.23  & \underline{34.39} & 16.14 & 21.97 & 45.27 & 73.65 & 17.10\\
    \multicolumn{1}{l|}{-w/o loss-based selection} & 18.54 & 34.09 & 12.34 & 1.24  & 26.87 & 18.65 & 22.02 & 45.55 & 73.96 & 17.10\\
    \multicolumn{1}{l|}{-w/o retrieval-based selection} & 18.71 & 34.91 & 12.21 & 1.26  & 32.33 & \underline{19.94} & 24.67 & 45.03  & 73.50 & 17.16\\
    \multicolumn{1}{l|}{-w/o data selection} & 18.27 & 34.44 & 12.03 & 1.21 & \textbf{34.71} & 16.42 & 22.29 & \underline{47.03} & \underline{74.64} & 16.65\\
    \midrule
    \multicolumn{1}{l|}{-w/o noise tolerant loss} & \textbf{18.93} & 34.96 & 12.26 & \textbf{1.29}  & 33.02 & 19.57 & \underline{24.58} & 46.64 & 74.46 & \textbf{16.56}\\
    \multicolumn{1}{l|}{-w/o consistency regularization} & 18.78 & \underline{35.03} & \textbf{12.40} & 1.27 & 33.82 & 19.29 & 24.57 & 46.81 & 74.55 & 16.70\\
    \bottomrule
    \end{tabular}
    \label{tab:ablation}}
\end{table*}

\textbf{Defect Detection.} Table~\ref{tab:defect} presents the results of defect detection. We can observe consistent improvement on overall performance as in the defect detection task: HINT\textsubscript{(5)} improves the F1 of three pre-trained models by  19.18\%, 13.79\%, and 16.50\%, respectively. 
This indicates that \tool can help pre-trained models to capture the patterns of vulnerable code snippets. Besides, by comparing the results of HINT\textsubscript{(1)} and HINT\textsubscript{(5)}, we can also observe that after multiple iterations,
\tool can achieve better performance, e.g., improving HINT\textsubscript{(1)} by 2.10\% F1 in average on UniXcoder.
 
\textbf{Assertion Generation.} For assertion generation, as shown in Table~\ref{tab:assert}, we can observe that \tool can improve all baseline pre-trained models by a large margin. On average, HINT\textsubscript{(1)} and HINT\textsubscript{(5)} improve the EM of these models by 11.09\% and 13.14\%, respectively. Specifically, on CodeBERT, HINT\textsubscript{(5)} improves its baseline by 21.24\%, 5.29\%, and 11.44\% in terms of EM, LCS, and ED, respectively. This indicates that the ability to better utilize the unlabeled data of \tool is also beneficial to generate accurate assertion statements. 
\begin{tcolorbox}[breakable,width=\linewidth,boxrule=0pt,top=1pt, bottom=1pt, left=1pt,right=1pt, colback=gray!20,colframe=gray!20]
\textbf{Answer to RQ1:} 
\tool consistently improves three pre-trained code models on all tasks and metrics, indicating its effectiveness in leveraging unlabeled data for the pre-trained code models.
\end{tcolorbox}
\vspace{-0.2cm}
 
\begin{table*}[t]
\centering
\caption{Experimental results on cross-domain scenario. ``*'' denotes statistical significance in comparison to the base models (i.e., two-sided $t$-test with $p$-value$<0.01$).}\label{tab:cross}
\aboverulesep=0ex
\belowrulesep=0ex
\scalebox{1}{
\begin{tabular}{l|cccc|cccc}
\toprule
\multirow{2}{*}{\textbf{Approach}} & \multicolumn{4}{c|}{Python $\to $ Java} & \multicolumn{4}{c}{Java $\to $ Python}\\
\cmidrule{2-9}
& BLEU-4 & ROUGE-L & METEOR & CIDEr & BLEU-4 & ROUGE-L & METEOR & CIDEr \\
\midrule
 {CodeBERT}  & 9.98 & 20.01 & 4.99 & 0.21 & 12.65 & 22.35 & 7.02 & 0.25\\
 {CodeBERT+\tool}  & \textbf{13.82}* & \textbf{27.71}* & \textbf{8.10}* & \textbf{0.60}* & \textbf{16.14}* & \textbf{30.33}* & \textbf{9.89}* & \textbf{0.58}* \\
\midrule
 {CodeT5}  & 7.75 & 12.55 & 7.12 & 0.29 & 14.81 & 30.59 & 9.66 & 0.84 \\
 {CodeT5+\tool}  & \textbf{14.09}* & \textbf{22.51}* & \textbf{9.28}* & \textbf{0.88}* & \textbf{16.85}* & \textbf{33.70}* & \textbf{10.77}* & \textbf{1.07}* \\
\midrule
 {UniXcoder}  & 12.68 & 26.03 & 8.33 & 0.66 & 13.18 & 20.94 & 10.70 & 0.64\\
 {UniXcoder+\tool}  & \textbf{16.33}* & \textbf{32.22}* & \textbf{10.54}* & \textbf{1.03}* & \textbf{17.26}* & \textbf{28.28}* & \textbf{13.14}* & \textbf{0.97}* \\
\bottomrule
\end{tabular}
}
\end{table*}


\subsection{RQ2: Ablation Study}\label{sec:RQ2}
In this section, we explore the contribution of the hybrid pseudo-labeled data selection and the noise-tolerant training modules proposed in \tool. 
We use UniXcoder as the base model since it shows the best performance in the first research question. Besides, considering the time and resource limitation of multiple iterations, we use \tool with one iteration for the following experiments.
Due to the page limit,
we only present the results on Java in this paper for code summarization, with results for other languages and pre-trained models presented
on our GitHub repository~\cite{HINT}. 

\subsubsection{Impact of hybrid pseudo-labeled data selection}
We compare \tool with four other data selection methods including \textit{Random selection}, \textit{\tool w/o retrieval-based selection}, \textit{\tool w/o loss-based selection}, and \textit{\tool w/o data selection}.
In \tool w/o loss-based selection and \tool w/o retrieval-based selection, we validate the effectiveness of two methods in hybrid pseudo-labeled data selection respectively. In \tool w/o data selection, we remove the whole data selection process and directly use all the generated pseudo-labeled data, which aims at verifying the benefit of data selection.
In Random selection, we randomly select a subset from pseudo-labeled data that has the same size as the subset selected by \tool. This is usually
used in controlled experiments to eliminate the potential confounding effect of dataset size~\cite{DBLP:journals/corr/abs-2207-05579}. The experimental results are presented in Table~\ref{tab:ablation}. 

\textbf{Loss-based selection.} We conduct this experiment by removing the loss-based selection (Line 4-5, 13-14 in Algorithm~\ref{alg:framework}). From Table~\ref{tab:ablation}, we can observe that, without loss-based selection, the performance of \tool decreases consistently
on all the tasks. Specifically, removing this component leads to an obvious decrease in defect detection, with the decrease at 19.26\%, 11.02\%, and 14.42\% regarding Precision, Recall, and F1, respectively. 
This demonstrates the benefits of removing the noisy data by the training loss.

\textbf{Retrieval-based selection.} We conduct this experiment by removing the retrieval-based selection (Line 9-11 in Algorithm~\ref{alg:framework}). As can be seen in Table~\ref{tab:ablation}, excluding the retrieval-based selection process leads to a consistent drop in all tasks and metrics. The results demonstrate the effectiveness of involving the retrieval-based strategy for data selection. 

\textbf{Data selection
and Random selection.} As shown in Table~\ref{tab:ablation}, the model suffers from a large degradation after removing the data selection procedure. Specifically, on defect detection, the F1 of using all pseudo-labeled data is only 22.29, much lower than the results of our method, i.e., 25.73. The performance of random selection is even worse. For example, on assertion generation, random selection has a decrease of 3.95\%, 1.43\%, and 2.95\% with respect to EM, LCS, and ED, respectively, indicating the importance of the data selection process in \tool.  
This also demonstrates that directly using pseudo-labeling cannot achieve promising results on code intelligence tasks.


\subsubsection{Impact of noise-tolerant training}
In this section, we validate the effectiveness of two components of the noise-tolerant training module, i.e., noise-tolerant loss function and consistency regularization.

\textbf{Noise-tolerant loss function.} We conduct this experiment by removing the noise tolerant loss in Equation~\ref{equ:robust}, i.e., directly using the cross entropy loss. From Table~\ref{tab:ablation}, we can observe that removing the noise-tolerant loss results in a performance decrease in the vast majority of cases. For example, on defect detection, \tool without the noise-tolerant loss
suffers from a decrease of 4.47\% in terms of F1. This shows the importance of using noise-tolerant loss to mitigate the negative impact of errors in pseudo labels on the model performance.

\textbf{Consistency regularization.} We conduct this experiment by removing the noise tolerant loss in Equation~\ref{equ:robust}, i.e., only use the first term in Equation~\ref{equ:robust}. As can be seen in Table~\ref{tab:ablation}, removing the adaptive regularization also leads to a drop in most tasks and metrics. Specifically, removing consistency regularization leads to a decrease of 0.63\%, 4.51\%, and 0.68\% on three tasks regarding BLEU-4, F1, and EM, respectively, indicating the effectiveness of providing reliable training objectives for leveraging the pseudo-labeled data.

 \begin{tcolorbox}[breakable,width=\linewidth,boxrule=0pt,top=1pt, bottom=1pt, left=1pt,right=1pt, colback=gray!20,colframe=gray!20]
 \textbf{Answer to RQ2:}
 All components in hybrid pseudo-labeled data selection module and noise-tolerant training module demonstrate 
 a positive effect on the performance of \tool.

 \end{tcolorbox}

\subsection{RQ3: Evaluation on Cross-domain Scenario}\label{sec:RQ3}
In some programming languages, there is often a shortage of training data.
For the data-limited scenarios, transfer learning is a popular solution which transfers
the knowledge of similar domains with sufficient data to the target domains~\cite{DBLP:conf/issta/LiXLX0022,DBLP:conf/sigsoft/WangYGP0L22}. In this section, we conduct experiments to study the effectiveness of \tool in cross-domain scenarios, 
in which the model is trained on the source domain and tested on the target domain with a different programming language.
We use the code summarization task for evaluation as it contains two kinds of programming language. Specifically, we first train a model on the Java/Python dataset as the source domain and then evaluate its performance on the test set of the other (Python/Java) dataset as the target domain. 
As shown in Table~\ref{tab:cross}, \tool can improve the performance of pre-trained code models in the cross-domain scenario by a large margin. {Specifically, \tool improves the BLEU-4 score of UniXcoder by 28.79\% and 30.96\% on Java to Python and Python to Java, respectively, indicating that \tool can effectively utilize the knowledge in those unlabeled data by pseudo-labeling. This also shows \tool's ability to enhance pre-trained code models in new programming languages,
regardless of any disparities in their data distributions.

 \begin{tcolorbox}[breakable,width=\linewidth,boxrule=0pt,top=1pt, bottom=1pt, left=1pt,right=1pt, colback=gray!20,colframe=gray!20]
 \textbf{Answer to RQ3:} 
\tool can substantially boost the performance of pre-trained code models in cross-domain scenarios where no annotated data exist in the target domain. 
 \end{tcolorbox}

\begin{figure}[t]
     \centering
     \begin{subfigure}[h]{0.23\textwidth}
        \centering
    	\includegraphics[width=1 \textwidth]{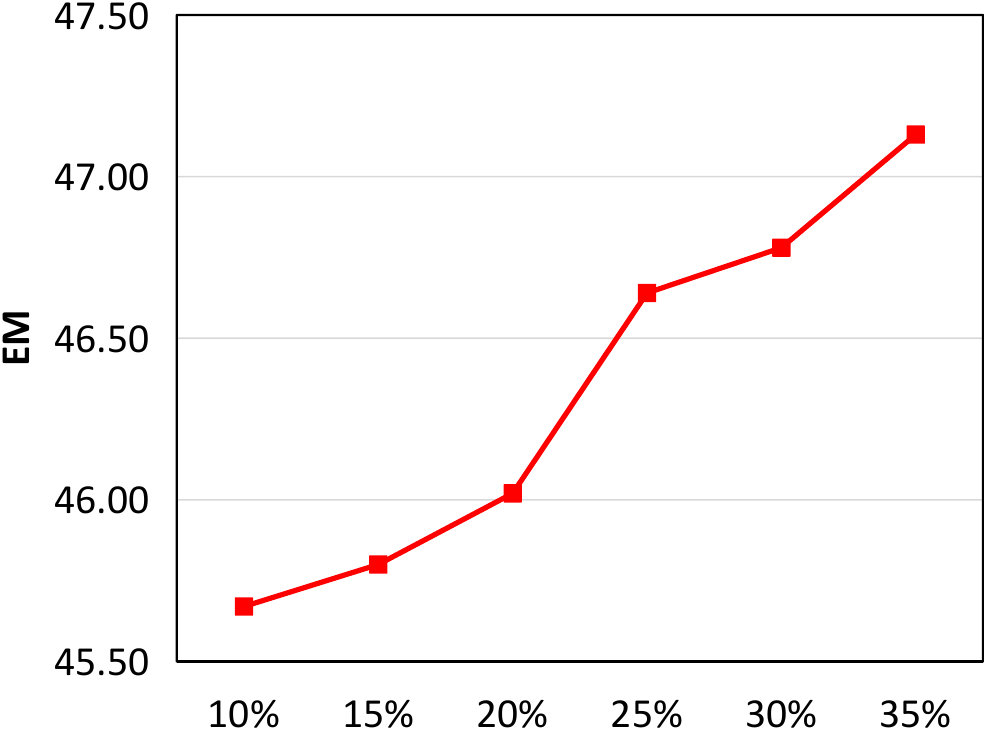}
    	\caption{\small Assertion generation.}
        \end{subfigure}
        \hfill
        \begin{subfigure}[h]{0.23\textwidth}
        \centering
        \includegraphics[width=1 \textwidth]{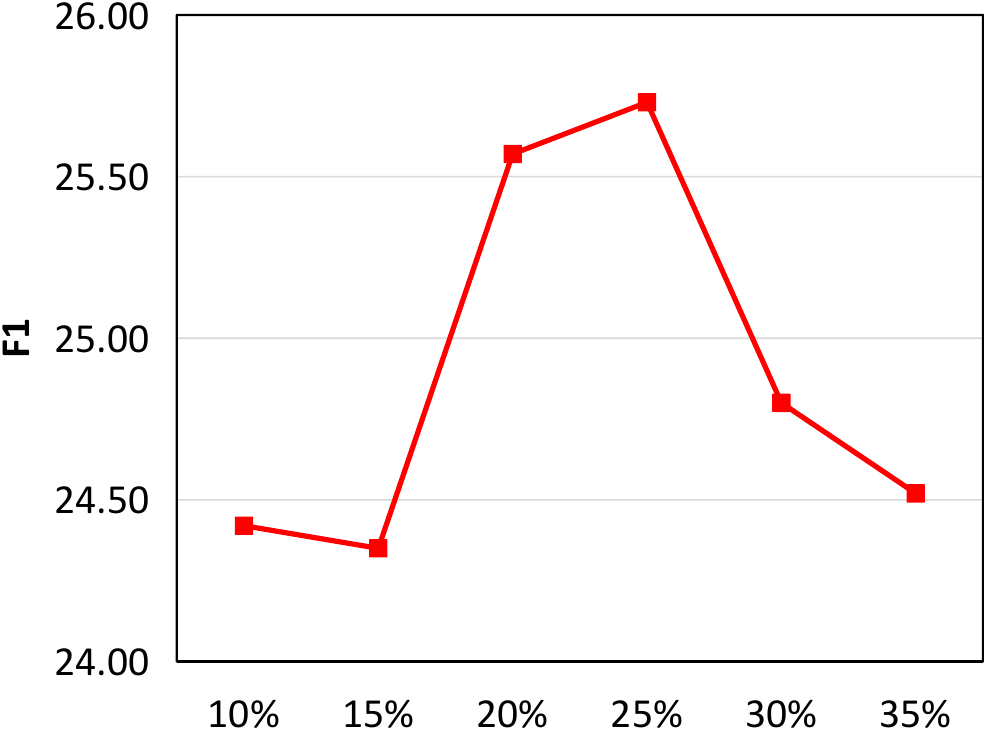}
        \caption{\small Defect detection.}
        \end{subfigure}
        \begin{subfigure}[h]{0.23\textwidth}
        \vspace*{0.15cm}
        \centering
        \includegraphics[width=1 \textwidth]{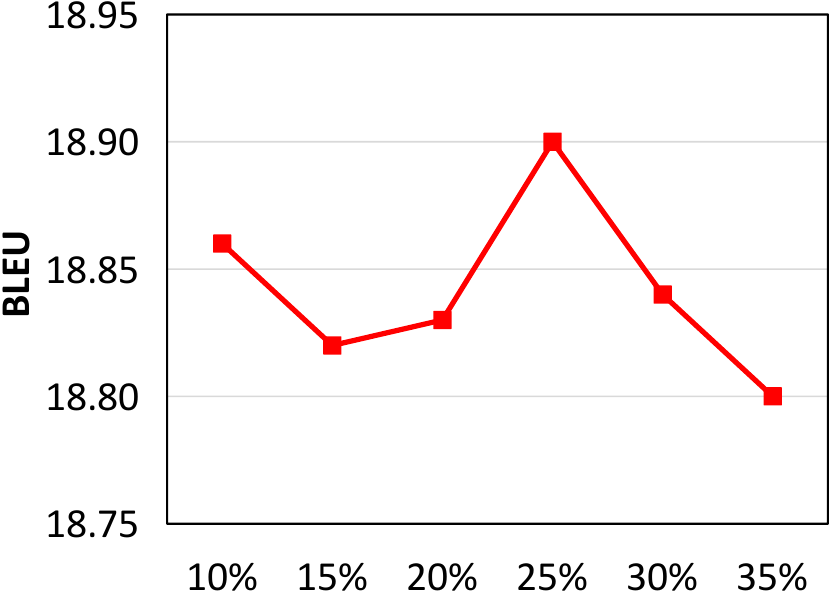}
        \caption{\small 
        Code summarization (UniXcoder).}
        \end{subfigure}
        \hfill
        \begin{subfigure}[h]{0.23\textwidth}
        \vspace*{0.15cm}
        \centering
        \includegraphics[width=1 \textwidth]{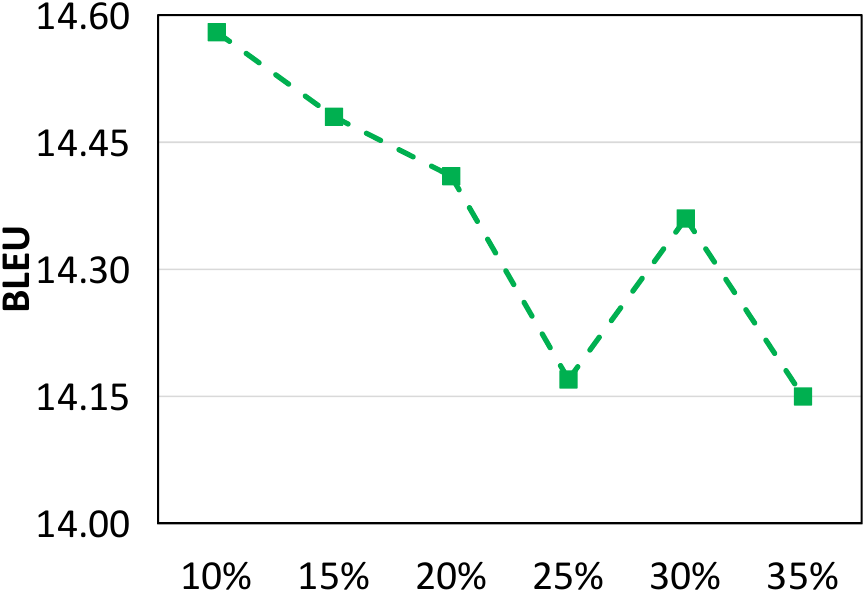}
        \caption{\small
        Code summarization (CodeBERT).}
        \end{subfigure}
        
         \caption{Parameter analysis on threshold $K$.}
     \label{fig:threshold}
\end{figure}

\begin{figure}[t]
     \centering
         \centering
         \includegraphics[width=0.47\textwidth]{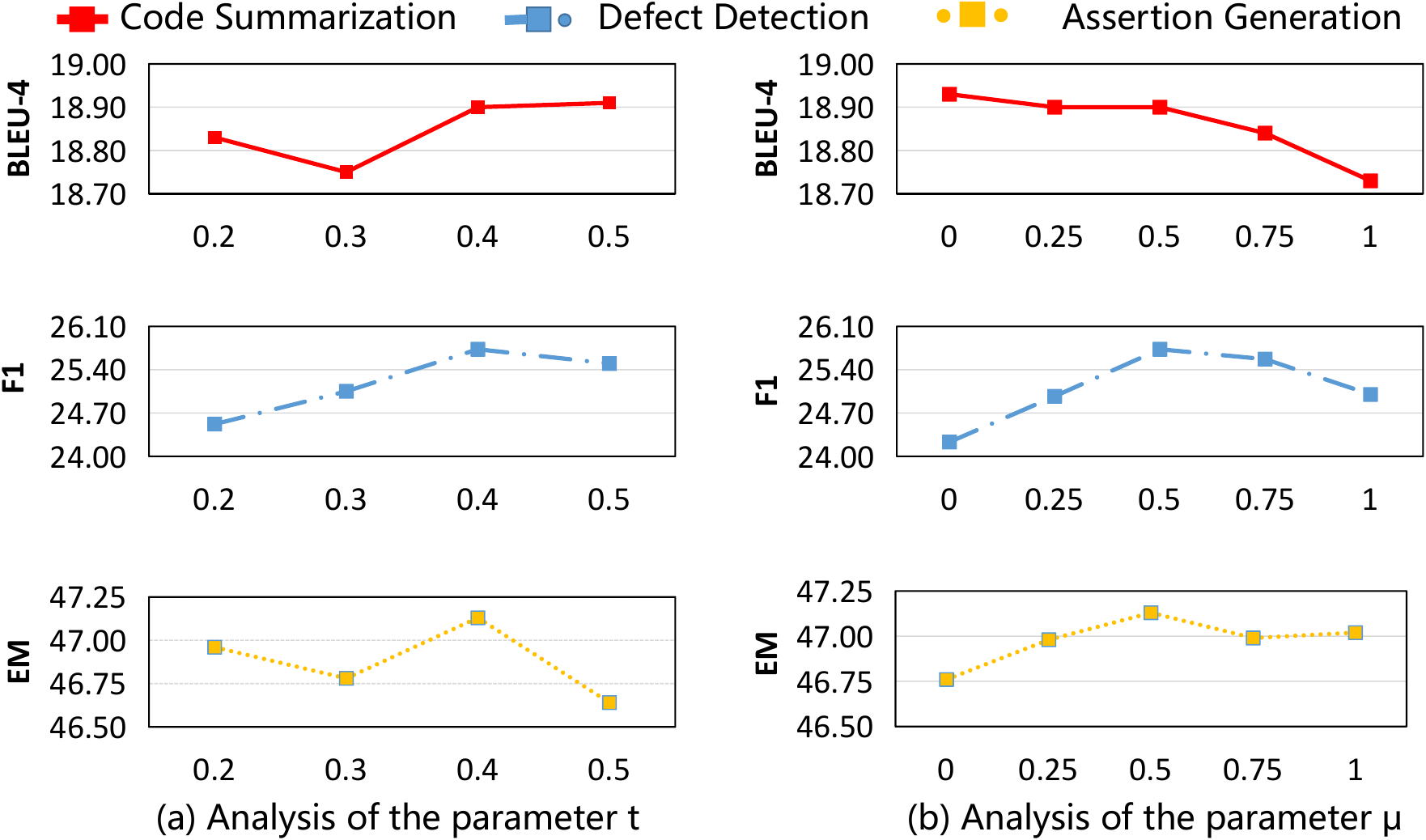}        
         \caption{Parameter analysis on $t$ and $\mu$.
         }
    \vspace{-0.4cm}
     \label{fig:parameter}
\end{figure}

\begin{figure}[t]
     \centering
     \begin{subfigure}[h]{0.23\textwidth}
        \centering
    	\includegraphics[width=1 \textwidth]{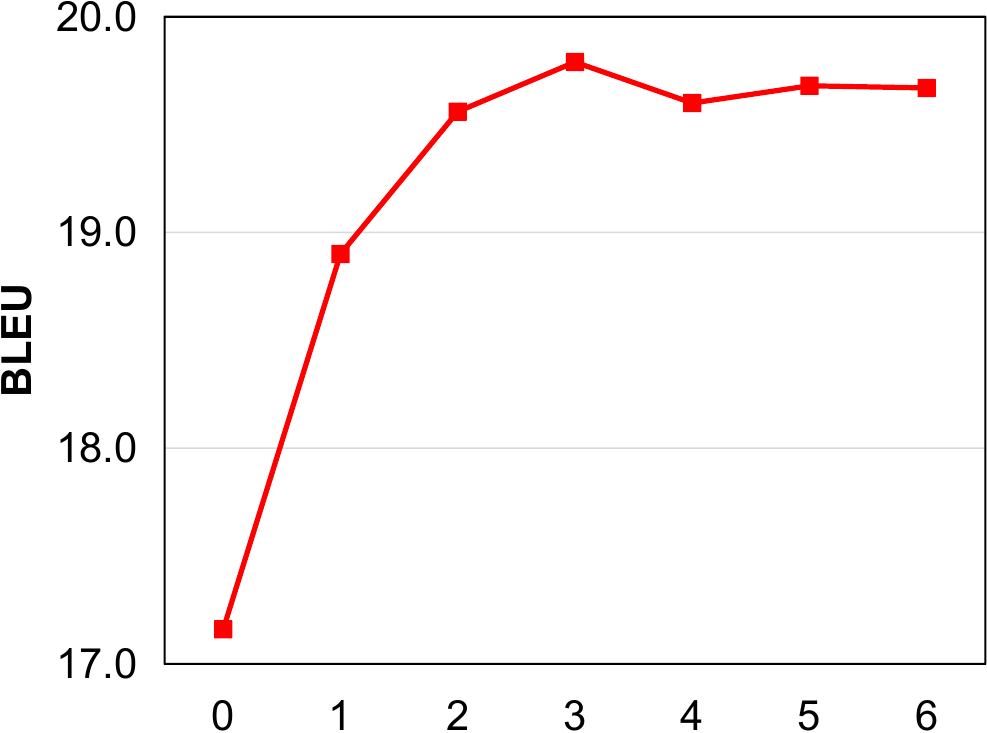}
    	\caption{Code sumamrization (JCSD).}
        \end{subfigure}
        \hfill
        \begin{subfigure}[h]{0.23\textwidth}
        \centering
        \includegraphics[width=1 \textwidth]{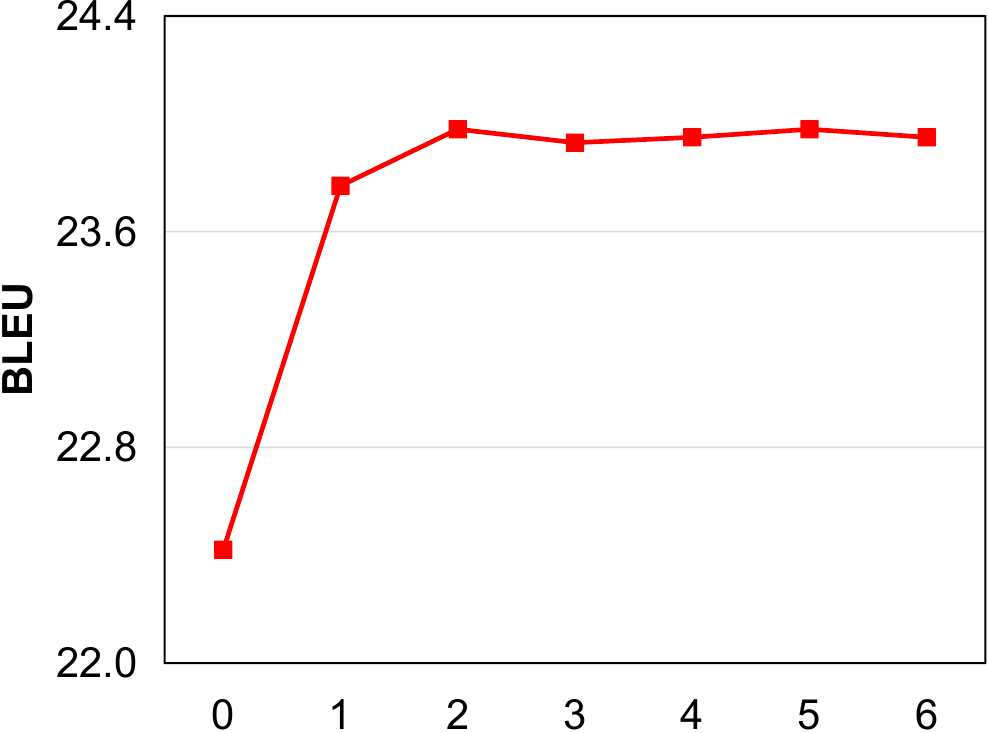}
        \caption{Code sumamrization (PCSD).}
        \end{subfigure}
        \begin{subfigure}[h]{0.23\textwidth}
        \vspace*{0.15cm}
        \centering
        \includegraphics[width=1 \textwidth]{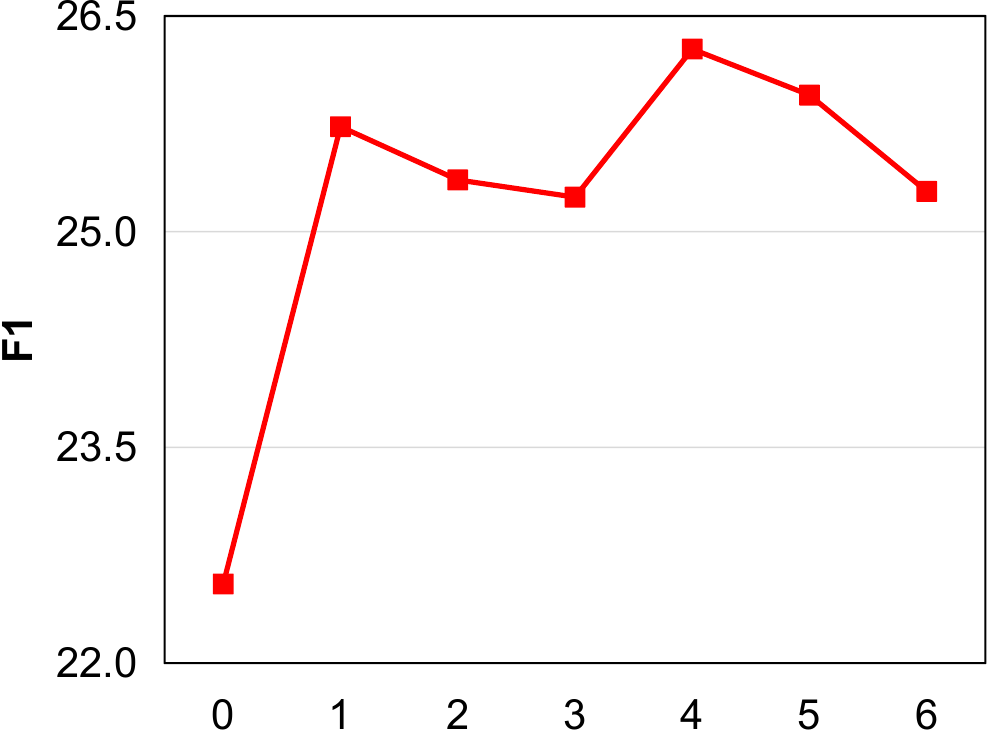}
        \caption{Defect detection.}
        \end{subfigure}
        \hfill
        \begin{subfigure}[h]{0.23\textwidth}
        \vspace*{0.15cm}
        \centering
        \includegraphics[width=1 \textwidth]{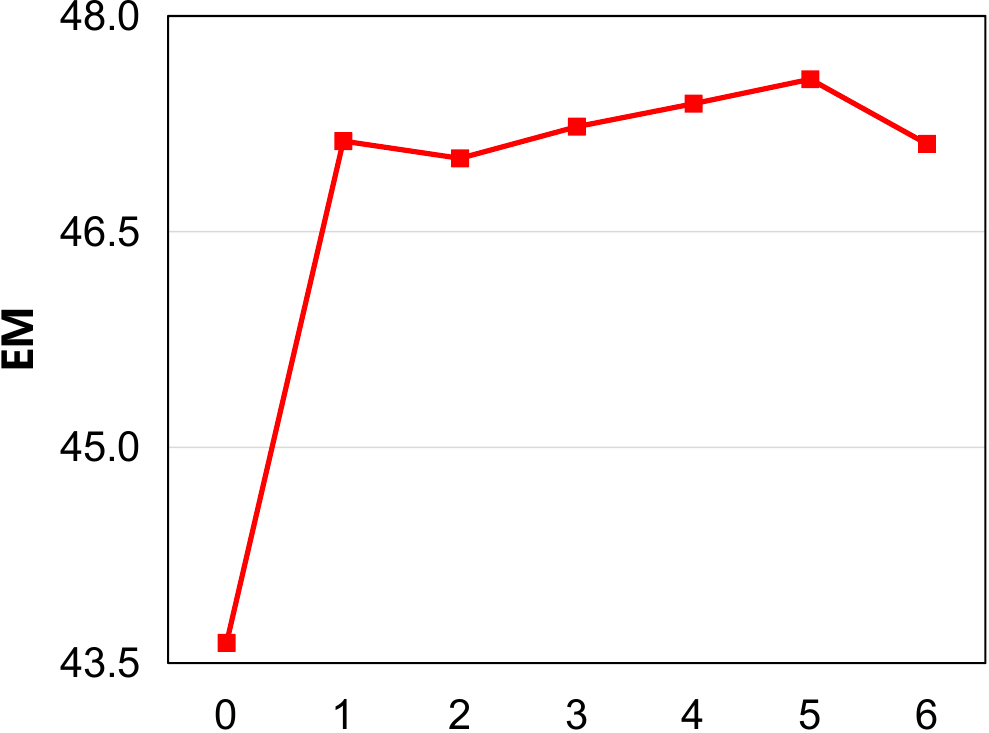}
        \caption{Assertion generation.}
        \end{subfigure}        
         \caption{Performance on each Iteration.} 
    \vspace{-0.4cm}
     \label{fig:trend}
\end{figure}

\subsection{RQ4: Parameter Analysis}
In this section, we study the impact of four parameters on the performance of \tool, including the threshold $K$ in loss-based data selection, the edit distance threshold $t$ in retrieval-based data selection, the weight of consistency regularization $\mu$, and the iteration number $I$. Due to the page limitation, we only present the results of
UniXcoder and JCSD dataset for $t$ and $\mu$ on code summarization, with results for other languages and pre-trained models presented
on our GitHub repository~\cite{HINT}. 

\textbf{The threshold $K$ in loss-based data selection.}
We conduct experiments to evaluate how \tool performs under different thresholds, i.e., 10\%, 15\%, 20\%, 25\%, 30\%, and 35\%. The larger the $K$ is set to, the more the pseudo labeled samples
will be selected. As shown in Figure~\ref{fig:threshold}, the model performance shows a similar trend along with the increase of $K$ on all pre-trained code models and tasks. \tool first increases and achieves its peak, and then sharply descends 
with a larger $K$. Larger $K$ has the risk of involving more noisy data while smaller $K$ might be too strict and filter many high-quality samples. Besides, we can find that the optimal value of $K$ for different models and tasks varies a lot. For example, on code summarization, UniXcoder achieves the best performance when $K$ is set to 25\%, while the optimal value for CodeBERT is 10\%. We suggest that it is because the capability of the base model on each task is different. Specifically, the performance of UniXcoder on code summarization is very strong, i.e. achieving 17.16 BLEU-4 on the Java dataset, while for CodeBERT, its performance on the Java dataset is only 13.30. The poorer the performance of the based model is, the lower the quality of pseudo-labeled data is.
Therefore, when applying \tool on different pre-trained code models, a relatively larger $K$ can be used on a stronger base model and vice versa.

\textbf{The edit distance threshold $t$.} We study the effect of $t$, as introduced in Section~\ref{subsec:aug}, by varying it from 0.2 to 0.5. As shown in Figure~\ref{fig:parameter} (a), for both defect detection and assertion generation, \tool achieves the best performance when $t$ is set to 0.4. Larger or lower values do not give better results. On code summarization, setting $t$ to 0.5 only performs slightly better than 0.4. This indicates that setting $t$ to 0.4 is more appropriate for \tool. Thus, we set $t$ to 0.4 in this work.

\textbf{The consistency regularization weight $\mu$.} To study the impact of $\mu$ in \tool, we vary it from 0 to 1 and show the results in Figure~\ref{fig:parameter} (b). Larger $\mu$ tends to give a stronger regularization to the model. For both defect detection and assertion generation, \tool achieves the best performance when $\mu$ is set to 0.5. 
However, on code summarization, increasing $\mu$ leads to a decrease in performance. Therefore, we set $\mu$ to 0.5 to enable \tool to produce relatively better results on different tasks.

\textbf{The iteration number $I$.} We evaluate the performance of \tool on different iterations by setting the maximum iteration to six, and present the results in Figure~\ref{fig:trend}. Iteration 0 represents the baseline results that do not use \tool. From the results, we can observe that \tool can get better results with the growth of iterations and achieves the peak at around the fifth iteration, indicating that \tool can achieve self-improvement by leveraging the unlabeled data.

 \begin{tcolorbox}[breakable,width=\linewidth,boxrule=0pt,top=1pt, bottom=1pt, left=1pt,right=1pt, colback=gray!20,colframe=gray!20]
 \textbf{Answer to RQ4:} 
Different settings of hyperparameters can influence the performance of \tool on different tasks. Our
hyper-parameter settings achieve relatively better results. 
 \end{tcolorbox}
\section{Discussion}\label{sec:discuss}

\subsection{What Makes \tool Work?}

\subsubsection{\tool can better utilize the unlabeled data for downstream tasks}
To better understand how pseudo-labeling benefits pre-trained code models, we give two examples in Figure~\ref{fig:case_sum} and Figure~\ref{fig:case_assert}. The case in Figure~\ref{fig:case_sum} shows a Java code snippet with summaries generated by UniXcoder and UniXcoder+\tool. From the example, we can see that the summary generated by UniXcoder only contains a simple description without a detailed introduction to the parameters. \tool can avoid this problem and give a more precise prediction since it can learn from more $\langle$unlabeled data, pseudo label$\rangle$ pairs that have a similar summary pattern. We also present another case in the assertion generation task in Figure~\ref{fig:case_assert}. The assertion statement generated by UniXcoder mistakenly predicts the assertion type as ``\textit{assertionTrue}'' since it does not learn the meaning of ``\textit{empty}'' well. However, since UniXcoder+\tool uses the code snippet in Figure~\ref{fig:case_assert} as training data which has the same assertion types as this test sample, it can correctly predict the assertion type in Figure~\ref{fig:case_assert}. 


\subsubsection{\tool can select pseudo-labeled data with higher quality}
Another advantage of \tool comes from our data selection process. \tool can select high-quality pseudo labels for model training. {As shown in Figure~\ref{fig:motivated_example} and \ref{fig:approach}, \tool identifies
low-quality pseudo-labeled data by employing both the implicit loss-based selection and explicit retrieval-based selection.} To further validate this, we calculate the edit distance 
of the pseudo labels generated by UniXcoder to the ground truth labels of the unlabeled dataset, and use the average distance on the whole selected dataset to measure the quality of our selected dataset. 
Specifically, on JCSD the average edit distance of all pseudo labels without filtering is 53.70, which is much higher than the dataset selected by \tool, i.e., 37.16. The results on assertion generation are the same. \tool achieves an average edit distance of 5.34 while the average edit distance of all pseudo labels is 17.86. 
This further shows that \tool can filter noisy data and select pseudo-labeled data with higher quality for model training.

\subsection{Limitation of HINT}
To gain a deeper understanding of HINT's behavior and limitations, we further investigate cases where HINT fails to make accurate predictions and conclude two possible limitations of HINT.

The first limitation pertains to HINT's inability to introduce additional knowledge and rectify factual knowledge errors. From the example in the above Figure~\ref{fig:case_fail1}, UniXcoder misinterprets the term ``\textit{bucket\_acl}'' as the name of a bucket and fails to rectify this misunderstanding even after additional training on pseudo-labeled data. This shows that without external feedback HINT is hard to identify and rectify the problem on factual knowledge, which also aligns with recent findings on the limited self-correction ability of large language models~\cite{DBLP:journals/corr/abs-2310-01798}. To potentially alleviate this limitation, integrating factual knowledge into pre-trained code models via the interaction with a knowledge base or search engine could be further studied.

The second limitation of HINT is the reliance on the capacity of the base model. HINT aims at autonomously synthesizing more labeled data for model training. However, when the base model lacks sufficient capacity, the benefits of additional training data are diminished. As depicted in the Figure~\ref{fig:case_fail2}, despite the presence of training sample in the pseudo-labeled data illustrating the usage of ``\textit{assertEquals}'', UniXcoder still fails to learn this and erroneously generates ``\textit{assertThat}'' for the given function. We attribute this limitation to the inherent constraints of the model's capacity and believe that it could be mitigated by using more advanced pre-trained code models.

 \begin{figure}
    \centering
    \includegraphics[width=0.45\textwidth]{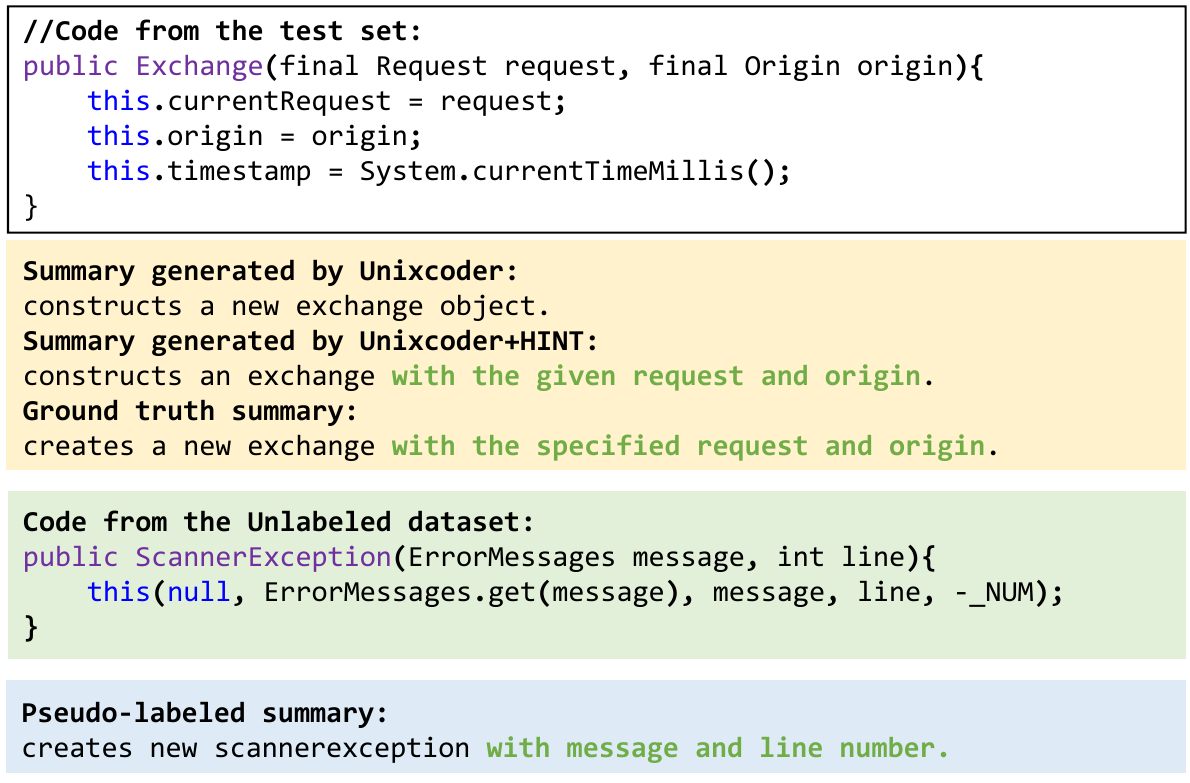}
    \caption{Case study on the code summarization task. The green texts highlight the similar part between the prediction of UniXcoder+\tool and pseudo-labeled summary.
    }
    \vspace{-0.2cm}
    \label{fig:case_sum}
\end{figure}

 \begin{figure}
    \centering
    \includegraphics[width=0.45\textwidth]{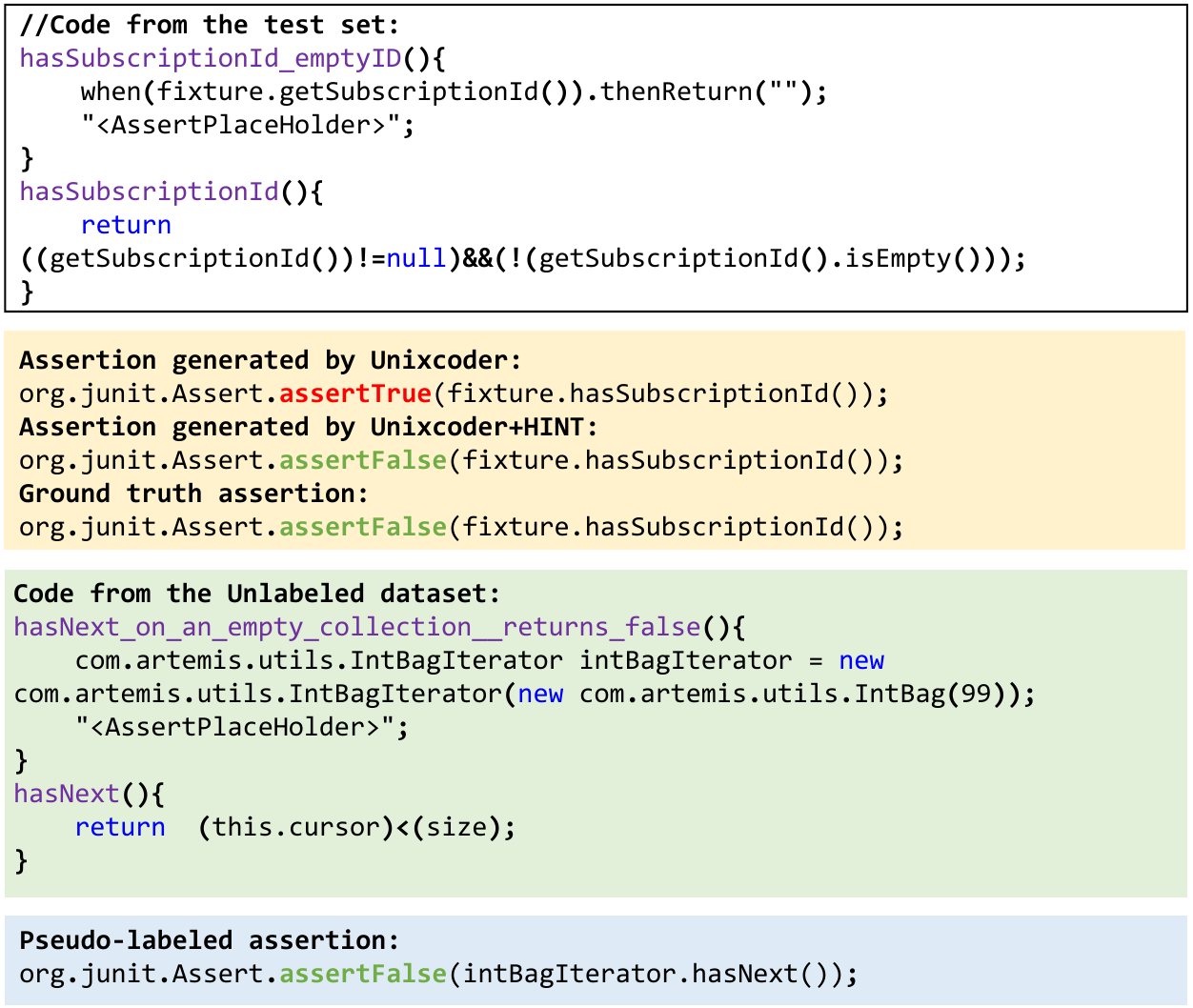}
    \caption{Case study on the assertion generation task. The red and green texts highlight the difference in predictions made by UniXcoder and UniXcoder+\tool.
    }
    \vspace{-0.2cm}
    \label{fig:case_assert}
\end{figure}

\subsection{Threats to Validity}

We identify  {four} 
main threats to validity of our study:

\begin{enumerate}
\item \textbf{The selection of code intelligence tasks.} We evaluate \tool on three commonly-used code intelligence tasks: code summarization, defect detection, and assertion generation. 
We aim to expand the validation of \tool in the future by testing it on more code intelligence tasks.

\item \textbf{The selection of pre-trained code models.} In this paper, we select three popular open-source pre-trained code models CodeBERT, CodeT5, and UniXcoder for evaluation. These models are all representative and have shown state-of-the-art performance on benchmarks~\cite{DBLP:conf/emnlp/0034WJH21,DBLP:conf/acl/GuoLDW0022}. 
Recent studies propose pre-trained models with much larger sizes such as ChatGPT~\cite{ChatGPT} and GPT-4~\cite{GPT4} which also show impressive programming ability. However, since the weight of these models is not publicly available, we cannot evaluate our framework on those large language models. Besides, our framework is flexible and easy to be applied to different pre-trained code models. 

\item \textbf{The selection of languages.} The datasets that we choose in experiments only contain two kinds of languages, i.e., Java and Python. They are both popular languages. Additionally, our method is language-agnostic and can be easily adapted to other programming languages. 

\item \textbf{The limitation of selected metrics.} {We evaluate \tool using a variety of commonly used metrics for different tasks. However, these metrics are mainly used for evaluating accuracy and may not reflect other evaluation
aspects such as the diversity of generated code summaries. In the future, we plan to conduct
human studies to provide a more comprehensive evaluation.}
\end{enumerate}


 \begin{figure}
    \centering
    \includegraphics[width=0.45\textwidth]{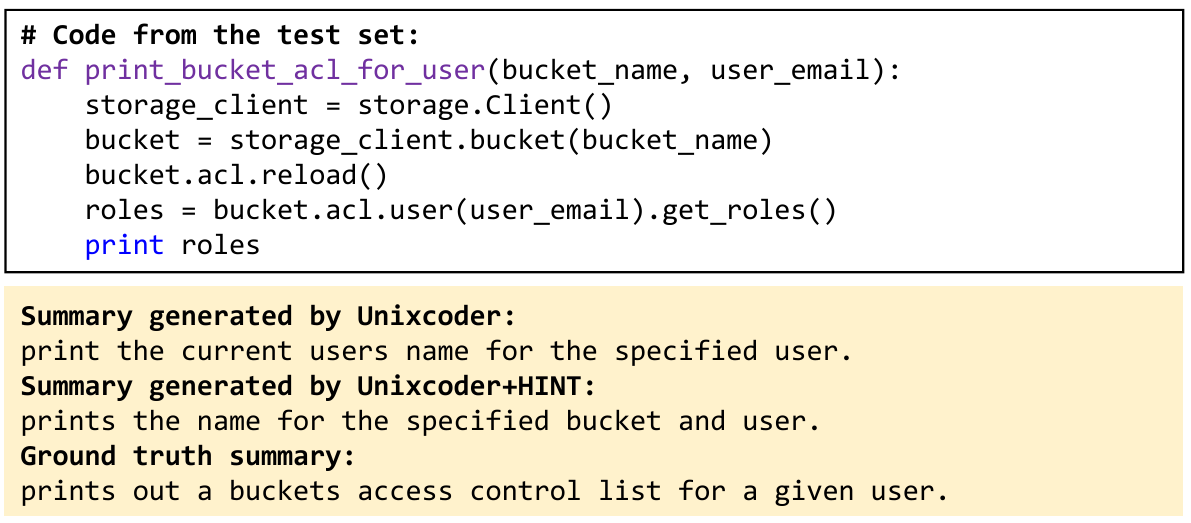}
    \caption{Error case on the code summarization task.
    }
    \vspace{-0.2cm}
    \label{fig:case_fail1}
\end{figure}

 \begin{figure}
    \centering
    \includegraphics[width=0.45\textwidth]{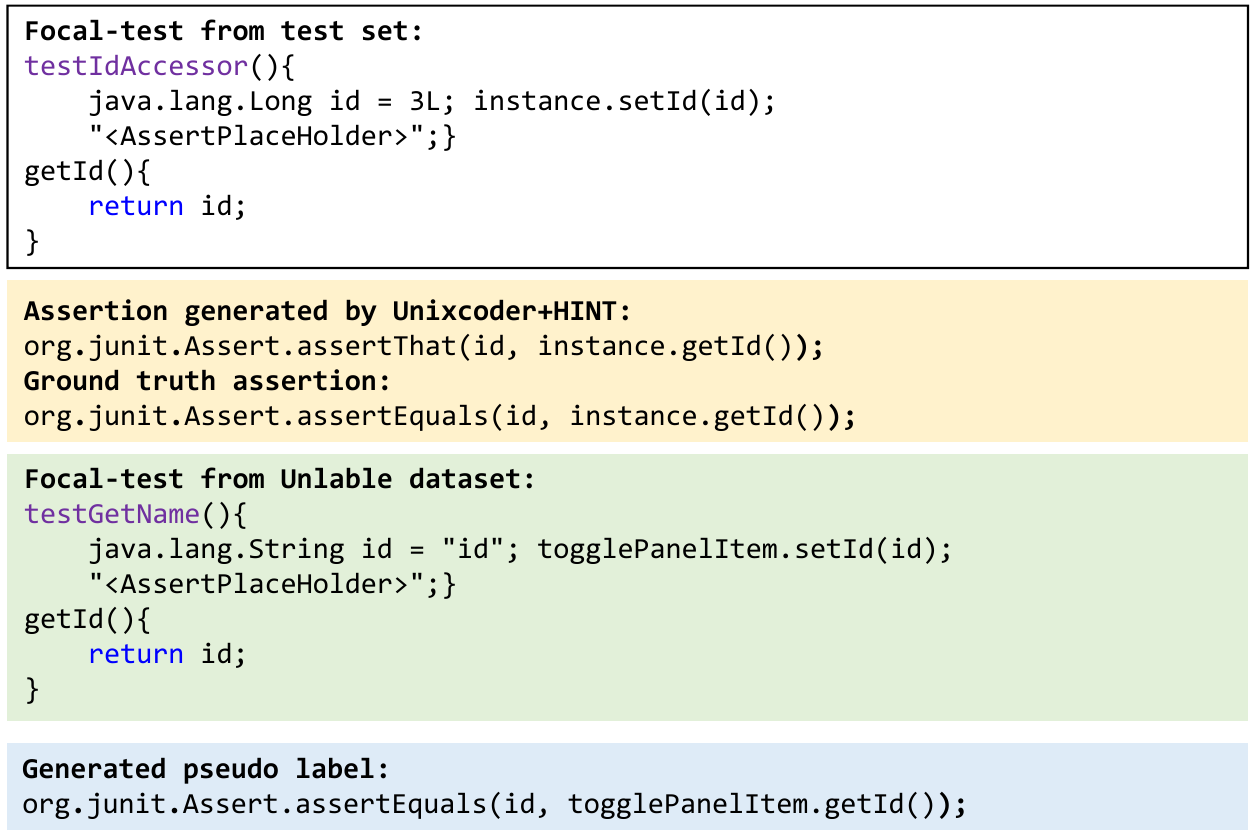}
    \caption{Error case on the assertion generation task. 
    }
    \vspace{-0.2cm}
    \label{fig:case_fail2}
\end{figure}
\section{Related work}\label{sec:related}



\subsection{Code Intelligence}
In this section, we introduce related neural code models in three tasks that are covered in our work, including both non-pre-trained code models and pre-trained code models,

\subsubsection{Non-pre-trained code models}
Iyer et al.~\cite{DBLP:conf/acl/IyerKCZ16} formulate code summarization as a neural machine translation (NMT) problem and propose CODE-NN to translate code snippets to code summaries. For better utilizing code structure information, many works~\cite{DBLP:conf/iclr/FernandesAB19,DBLP:conf/iwpc/LeClairHWM20} in code summarization also incorporate code-related graphs and GNN to boost performance. Recent studies~\cite{DBLP:conf/acl/WuZZ21,DBLP:conf/icse/TangSLGHZ022} further incorporate various code structure information into the Transformer model and achieve promising performance.  As for vulnerability detection, many deep learning-based methods~\cite{DBLP:conf/sigsoft/Li0N21,DBLP:conf/nips/ZhouLSD019} are proposed. For example, Devign~\cite{DBLP:conf/nips/ZhouLSD019} is proposed to learn the various vulnerability characteristics with a composite code property graph and graph neural network.   IVDetect~\cite{DBLP:conf/sigsoft/Li0N21} uses the program dependency graph and feature attention GCN to detect vulnerabilities in the code. In assertion generation, recent studies adopt the T5 transformer model and achieve promising results~\cite{mastropaolo2021studying,mastropaolo2022using}. Yu et al.~\cite{DBLP:conf/icse/YuLSR00L0W22} further involve information retrieval to generate more accurate assertion statements.

\subsubsection{Pre-trained code models}
Recently, a series of pre-trained code models~\cite{DBLP:conf/emnlp/FengGTDFGS0LJZ20,DBLP:conf/iclr/GuoRLFT0ZDSFTDC21,DBLP:conf/emnlp/0034WJH21} are proposed and achieve state-of-the-art performance on various code intelligence tasks such as code summarization and defect detection. CodeBERT~\cite{DBLP:conf/emnlp/FengGTDFGS0LJZ20} is a pioneer work that is pre-trained with six programming languages and uses Masked Language Modeling and Replace Token Detection as pre-trained tasks. 
CodeT5~\cite{DBLP:conf/emnlp/0034WJH21} is a sequence-to-sequence pre-trained model which involves two code-related pre-training objectives: identifier tagging and masked identifier prediction. 
UniXcoder~\cite{DBLP:conf/acl/GuoLDW0022} is a unified cross-modal pre-trained model which incorporates code semantic and syntax information from AST. 

\subsection{Pseudo-labeling}
Pseudo-labeling is one of the most widely-used semi-supervised learning methods. 
It has been applied to different kinds of tasks such as image classification~\cite{DBLP:conf/cvpr/XieLHL20,lee2013pseudo}, machine translation~\cite{DBLP:conf/acl/Jiao0T0LK20,DBLP:conf/iclr/HeGSR20}, and dialog systems~\cite{DBLP:conf/emnlp/MiZ0CHF21}. To further boost the performance of self-training in sequence generation tasks, He et al.~\cite{DBLP:conf/iclr/HeGSR20} and Mi et al.~\cite{DBLP:conf/emnlp/MiZ0CHF21} explore the data augmentation technique and use random noise or gradient-based data augmentation to improve the generalization of the student model. Another line of work~\cite{DBLP:conf/acl/Jiao0T0LK20,DBLP:conf/coling/WangWWC22,DBLP:conf/emnlp/ChenZZLC021} focus on the data selection procedure and propose to select high-quality pseudo labeled data based on the uncertainty or the model confidence, respectively. 
However, these methods mainly filter the pseudo-labeled data only with training loss and do not take the noisy data problem into consideration. Different from them, we propose a hybrid data selection method with the training loss and a retrieval-based method based on the code reuse. Additionally, we also propose a noise-tolerant training module to further mitigate the influence of noise on model performance. 

\section{Conclusion}\label{sec:conclusion}
In this paper, we investigate leveraging large-scale unlabeled datasets for effectively tuning pre-trained code models by pseudo-labeling.
We propose a method called \tool which consists of two main components, the hybrid pseudo-labeled data selection module and the noise-tolerant training module. Extensive experiments on three code intelligence tasks show that \tool can be built on a variety of pre-trained models and provide complementary benefits for them. Our replication package including our source code, experimental data, and detailed experiment results is at~\cite{HINT}. 




\bibliographystyle{ACM-Reference-Format}
\bibliography{sample}
\end{document}